\documentclass[sigplan,10pt,nonacm]{acmart}

\newcommand{\cwu}[1]{\textcolor{red}{[CWu: #1]}}
\newcommand{\jt}[1]{\textcolor{blue}{[Jt: #1]}}
\newcommand{\ljr}[1]{\textcolor{teal}{[LJr: #1]}}
\newcommand{\sys}{Sandwich}

\newtheorem{assumption}{Assumption}

\usepackage{amsfonts}
\usepackage{comment}
\newtheorem{proposition}{Proposition}

\newcommand{\TG}{\mathcal{S}}
\newcommand{\bigO}{O}
\newcommand{\BigF}{\mathcal{F}}
\newcommand{\bigF}{F}
\newcommand{\smlF}{f}

\newcommand{\gmk}{\mathtt{gemm\_micro\_kernel}}
\newcommand{\geff}{\mathtt{gemm\_e}}
\newcommand{\mkeff}{\mathtt{micro\_kernel\_efficency}}

\usepackage{tikz}
\newcommand*\circled[1]{\tikz[baseline=(char.base)]{
            \node[shape=circle,draw,inner sep=1pt] (char) {#1};}}

\usepackage{subcaption}
\usepackage{multirow}
\usepackage{algpseudocode}

\usepackage{algorithm}

\usepackage{listings}
\definecolor{dkgreen}{rgb}{0,0.6,0}
\definecolor{gray}{rgb}{0.5,0.5,0.5}
\definecolor{mauve}{rgb}{0.58,0,0.82}
\newcounter{observationcount}
\newcommand{\observation}[1]{
  \stepcounter{observationcount}
  \noindent
  \textbf{Observation \#\theobservationcount: #1}
}

\AtBeginDocument{%
  }

\setcopyright{none} 
\copyrightyear{}
\acmYear{}
\acmDOI{}

\acmISBN{}

\renewcommand\footnotetextcopyrightpermission[1]{} 
\author{Juntao Zhao$^{\dagger}$}
\affiliation{%
  \institution{The University of Hong Kong}
  \city{Hong Kong}
  \country{China}
}
\email{jtzhao@connect.hku.hk}

\author{Jiuru Li$^{\dagger}$}
\affiliation{%
  \institution{The University of Hong Kong}
  \city{Hong Kong}
  \country{China}
}
\email{u3011183@connect.hku.hk}

\author{Chuan Wu}
\affiliation{%
  \institution{The University of Hong Kong}
  \city{Hong Kong}
  \country{China}
}
\email{cwu@cs.hku.hk}
\begin{document}

\title{Sandwich: Joint Configuration Search and Hot-Switching for
Efficient CPU LLM Serving}




\begin{abstract}

CPUs are critical for LLM serving due to their availability, cost efficiency, and edge applicability. However, efficient CPU serving is hindered by conflicting prefill/decode resource demands under non-disaggregated deployment constraints—existing solutions fail to avoid cross-phase interference, ignore sub-NUMA hardware structures, and deliver suboptimal dynamic-shape kernel performance.
We propose \sys{}, a full-stack CPU LLM serving system with three core innovations addressing these challenges: (1) seamless phase-wise plan switching to eliminate cross-phase interference; (2) TopoTree, a tree-based hardware abstraction for automated substructure-aware (e.g., LLC slices) partial core allocation; (3) fast-start-then-finetune dynamic-shape tensor program generation.
Across five x86/ARM CPU platforms, \sys{} achieves an average 2.01× end-to-end speedup and up to 3.40× latency reduction over state-of-the-art systems. Its kernels match static compiler performance with three orders of magnitude lower tuning cost.

\end{abstract}



\maketitle
{\let\thefootnote\relax\footnote{{$^\dagger$Equal contribution.}}}

\section{Introduction}\label{sec:intro}
The scarcity of high-end GPUs~\cite{zhao2024llmpq}, combined with the demand for greater availability and flexibility in data centers~\cite{character, facebook}, has increased interest in using CPUs for language model deployment, owing to their widespread availability and cost efficiency. The emergence of capable small- and medium-scale language models (SLMs)~\cite{slm_future_agent} for specialized tasks, along with power-constrained edge environments~\cite{small_scale}, further solidifies the CPU’s role as a vital deployment platform.

%
Efficient CPU-based LLM serving faces two primary challenges. As prior work \cite{zhong2024distserve} establishes, LLM inference consists of two phases with divergent resource demands: a compute-intensive \emph{prefill} phase that processes variable-length inputs, and a memory-intensive \emph{decode} phase that is bottlenecked by memory bandwidth and cache contention. First, strict memory constraints in many CPU deployment scenarios (e.g., edge devices) mandate \textbf{non-disaggregated} generation: prefill and decode must be colocated within a single instance to avoid redundant model weight storage, which amplifies the need to unify optimizations for their conflicting resource requirements. At the same time, modern server CPUs exhibit complex hierarchical memory architectures featuring Non-Uniform Memory Access (NUMA) and shared cache hierarchies (e.g., multiple Last-Level caches (LLCs)). Thus, low-latency efficient serving requires four key capabilities: (1) optimize variable-length prefill tensor programs, (2) alleviate decode’s memory bottlenecks, (3) align optimizations with underlying CPU hardware, and (4) minimize prefill-decode cross-phase interference in colocated instances.

\begin{figure}
\centering
\includegraphics[width=\linewidth]{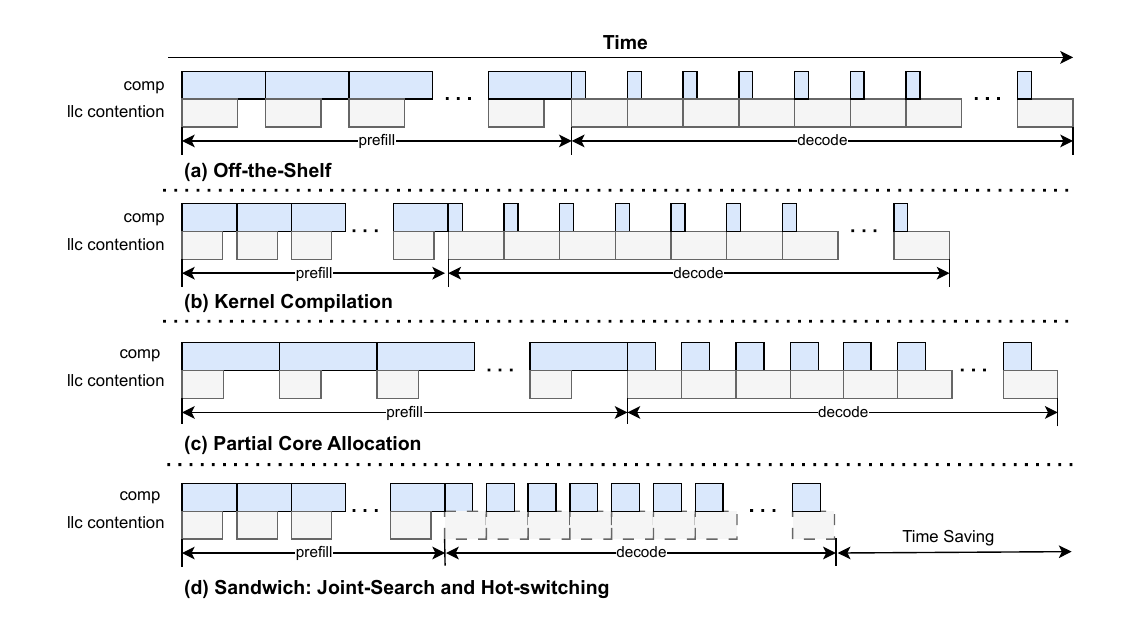}
\caption{Execution time comparison of CPU serving solutions. The prefill stage benefits from kernel compilation due to higher kernel efficiency, while the decode stage improves with partial core allocation, which reduces last-level cache (LLC) contention. By dynamically switching computation graphs, \sys{} avoids interference from partial core allocation, enabling fine-grained control and better performance (dashed line) than either approach alone.}
\label{fig:three_sols}
\end{figure}

Existing systems like CPU backend of vLLM~\cite{vllm}, llama.cpp~\cite{llama.cpp}, and xFasterTransformer~\cite{xFasterTransformer}—often rely on vendor-specific or hand-tuned solutions with limited adaptability to dynamic shapes. As shown in Fig.~\ref{fig:three_sols}(b), automatic tensor program schedulers like TVM~\cite{TVM} can generate dynamic-shape results, but require substantial auto-tuning effort. To reduce search overhead, recent dynamic-shape compilers~\cite{Roller, milkpoly, dietcode} encode hardware primitives into micro-kernels (MKs), which are programs that partially compute results, and apply \emph{polymerization schemes}, which define a selection of MKs to use and how they are assigned to computing resources under different input shapes.

On the other hand, since CPUs offer fine-grained core control via APIs like OpenMP~\cite{openmp08}, several serving systems~\cite{vllm} advocate partial core allocation to limit parallelism and reduce memory contention during the memory-bound decode phase~\cite{chen2023punica}, as illustrated in Fig.~\ref{fig:three_sols}(c). To avoid cross-NUMA memory accesses~\cite{manycoreInf}, many systems employ NUMA-aware thread scheduling~\cite{arm_manycore_trans, llama.cpp, openvino} or partition models evenly across NUMA nodes~\cite{xFasterTransformer}.

Despite these optimizations, current approaches remain suboptimal for CPU-based serving. For the decode phase, partial core allocation reduces prefill parallelism, hurting its performance. Moreover, the derived core allocation schemes often ignore sub-NUMA structures, such as the four-core clusters sharing an LLC slice in AMD EPYC 7H11 processors~\cite{amd_server_processor}. For the polymerization scheme at prefill, dynamic-shape compilers either adopt a scale-up-and-out strategy~\cite{Roller}, i.e., expanding MK shapes greedily per cache level and partitioning across processors, or use a cost-model-based approach with fixed-size MKs and runtime polymerization. While larger MKs improve kernel efficiency, they reduce parallelizability, constraining the polymerization search space. Existing methods fail to model this trade-off, leading to suboptimal performance. 

To address these challenges, we propose \sys{}, a full-stack LLM serving system for CPUs that enables program hot-switching between prefill and decode phases and efficiently explores the combinatorial design space of core allocations and dynamic-shape inputs. As illustrated in Fig.~\ref{fig:three_sols}(d), for core allocation, \sys{} represents CPU topology as a tree (TopoTree), systematically enumerating core allocation plans that account for shared hardware resources like LLC slices. For tensor program generation, \sys{} produces multiple MK shapes and explores polymerization schemes via a fast-start-then-finetune strategy, jointly optimizing computation slices and parallelization plans.  Our contributions are:

$\triangleright$ We design and implement a runtime hot-switching mechanism for CPU-based LLM serving, enabling separate execution plans for prefill and decode phases without requiring duplicate model copies.

$\triangleright$ We introduce TopoTree, a tree-based hardware abstraction that uses grouping and removal transformations to efficiently explore core allocation strategies, maximizing synergy and minimizing resource contention.

$\triangleright$ We develop a fast-start-then-finetune method for generating dynamic-shape tensor programs, which jointly optimizes MKs and polymerization schemes. This reduces tuning overhead and improves prefill kernel performance.

$\triangleright$ We evaluate \sys{} extensively on multiple CPU platforms (Xeon Gold 6151, 6230, Platinum 8272CL, EPYC 7H12, and Kunpeng 920) using diverse chatbot traces. \sys{} reduces latency by up to $3.40\times$ and achieves an average $2.01\times$ end-to-end speedup over the best existing systems, while matching TVM’s kernel performance with three orders of magnitude less tuning cost.

\section{Background}\label{sec:background}

\begin{figure}[t]
    \centering
    \begin{subfigure}[b]{0.43\linewidth}
        \centering
        \includegraphics[width=\linewidth, height=1.15in]{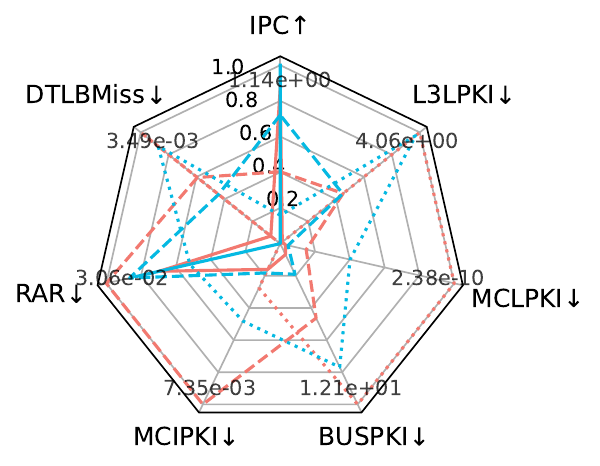}
        \caption{Prefill}
        \label{fig:ch-workload-prefill}
    \end{subfigure}
    \hfill 
    \begin{subfigure}[b]{0.51\linewidth}
        \centering
        \includegraphics[width=\linewidth, height=1.15in]{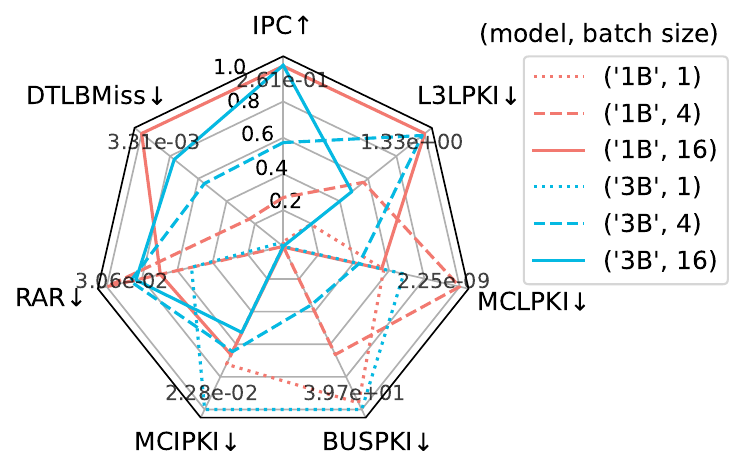}
        \caption{Decode}
        \label{fig:ch-workload-decode}
    \end{subfigure}

    \caption{Prefill and decode workload characterization using different performance counters for serving Llama3.2-1B and Llama3.2-3B.}
    \label{fig:ch-workload}
\end{figure}

\subsection{Prefill-decode workload difference in CPU}\label{sec:prefill_decode_trade_off}
To quantify the distinct workload characteristics of prefill and decode phases on many-core CPUs, we adopt performance metrics from~\cite{wasp, holistic_numa_memory}: Instructions Per Cycle (IPC), Last-level Cache Loads Per Thousand Instructions (L3LPKI), Data TLB Miss Ratio (DTLBMiss), Remote Access Ratio (RAR), Memory Controller Loads Per Thousand Instructions (MCLPKI), Memory Controller Imbalance (MCIPKI), and Bus Cycles Per Thousand Instructions (BUSPKI).

We evaluate BF16 Llama3.2-1B and 3B models on a dual-socket Intel Xeon Platinum 8275CL system (24 cores per socket, two NUMA nodes) using vLLM with NUMA-aware model partitioning. Requests are sampled from the ShareGPT dataset with batch sizes of 1, 4, and 16. As shown in Fig.~\ref{fig:ch-workload}, the prefill phase exhibits higher IPC and lower BUSPKI, indicating a compute-bound workload where memory accesses are spaced apart by computation, reducing pressure on memory controllers. In contrast, the decode phase shows lower IPC with significantly higher BUSPKI and MCLPKI, reflecting memory-bound behavior. Thus, prefill benefits from greater computational throughput, while decode requires memory contention mitigation.

\subsection{CPU Autonomy}
CPUs provide fine-grained core management through mechanisms like OpenMP, enabling precise control over core binding and affinity. As demonstrated in Sec.~\ref{sec:prefill_decode_trade_off}, memory bus congestion can be mitigated by strategically deactivating CPU cores, which improves end-to-end token generation throughput while reducing power consumption. Existing systems such as vLLM typically address this by evenly distributing models and computation across NUMA nodes, combined with limited core deactivation (e.g., 2 cores) to enhance overall efficiency.

\subsection{Dynamic-shape Tensor Program Generation}
Static-shape compilers~\cite{TVM} require unacceptable tuning times to produce dynamic shape tensor programs, because they need to tune every input shape. Dynamic-shape compilers construct tensor programs using \emph{micro kernels}, which are single-threaded programs that cover certain tensor shapes, and \emph{polymerization schemes}, which are schedules to place the computation slices onto parallel computation resources. They can be categorized into two approaches. 

Scale-up-and-out \cite{Roller} is a bottom-up approach that gradually increases the size of the current MK by repeating a smaller MK along some dimension and selecting the scale-up dimension that yields the most performance gain through profiling. When the performance improvement saturates at a cache level, the same process is repeated on the new MK. Eventually, a tiled MK is constructed for the LLC and it is replicated among CPU cores. A cost-model-based method \cite{dietcode, milkpoly} is a top-down approach that uses fixed-shape pre-compiled MKs, whose polymerization schedules are determined at runtime by a cost model. Our observation shows that both approaches fail at providing the optimal tensor program. 

\section{
\sys{} Overview}
\begin{figure}[t]
    \centering
    \vspace{-2mm}\includegraphics[width=\linewidth]{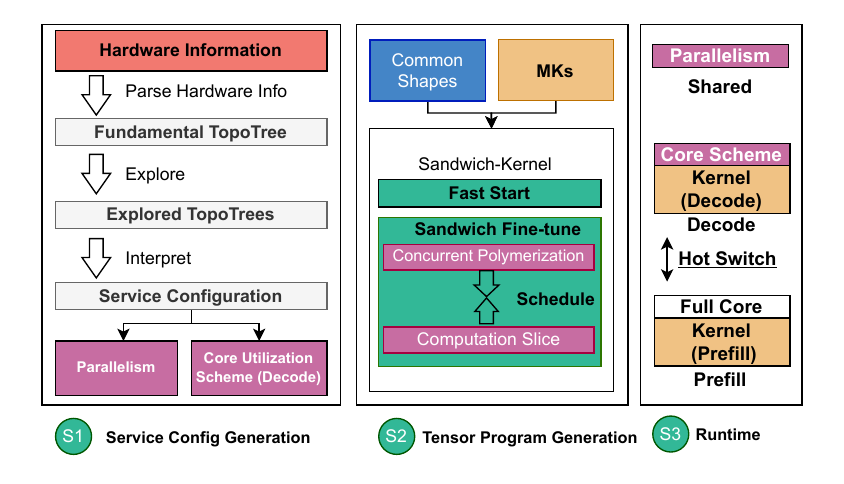}
    \Description{Workflow of Sandwich: In Sandwich, service configuration generation and kernel orchestration are performed offline. It begins by using hardware information to generate the foundational TopoTree. This tree is then explored through group transformations to uncover latent shared structures and resources. Once new candidate TopoTrees are generated, a remove transformation is applied to mitigate potential contention along the memory hierarchy. This transformation forms a transformation tree, which eliminates redundancies and enables efficient pruning.
Afterward, the TopoTrees are converted into service configurations, including process and thread spaces. By introducing a horizontal cross-section along the tree depth, we generate all service configurations that correspond to the tree. Several candidate service configurations (top-k) are then selected for kernel generation.
In the kernel generation phase, a sampled trace from the user side is taken to automatically generate payloads based on the model configuration. The kernel performance is then tuned by replaying the trace, simulating actual performance. During kernel orchestration, candidate micro-kernels (MKs) are generated using hardware information and common shapes provided by the payload generator. These candidate MKs are orchestrated with a fast-start and fine-tuning method to find the optimal combination of computation slices and concurrent polymerization.
Once the service configurations and tuned kernels are generated, they can be directly deployed by the Sandwich runtime in real LLM serving applications.}
    \caption{Workflow of Sandwich. 
    } 
    \label{fig:wf_sandwich}
\end{figure}


\sys{} comprises two core components, as illustrated in Fig.~\ref{fig:wf_sandwich}: service configuration generation (S1) and kernel orchestration (S2), both executed offline.
For service configuration generation, \sys{} first parses the CPU (NUMA) architecture via system tools to construct the fundamental \textit{TopoTree}. An exploration phase then enumerates all potential latent sub-structures among tree nodes using group transforms; it further employs remove transforms to discover all valid core utilization plans—wherein a subset of cores is reserved for decode operations. The resulting TopoTree is translated into two key schemes: a shared parallelism scheme and a core utilization scheme tailored for the decode phase.
For tensor program generation, \sys{} produces two sets of dedicated kernels: prefill kernels aligned with the prefill service configuration, and decode kernels matched to the decode service configuration. At runtime, these paired configurations and their corresponding optimized kernels are executed via hot-switching (S3).

\begin{figure*}[t]
    \centering
    \includegraphics[width=\linewidth, height=0.16\textheight]{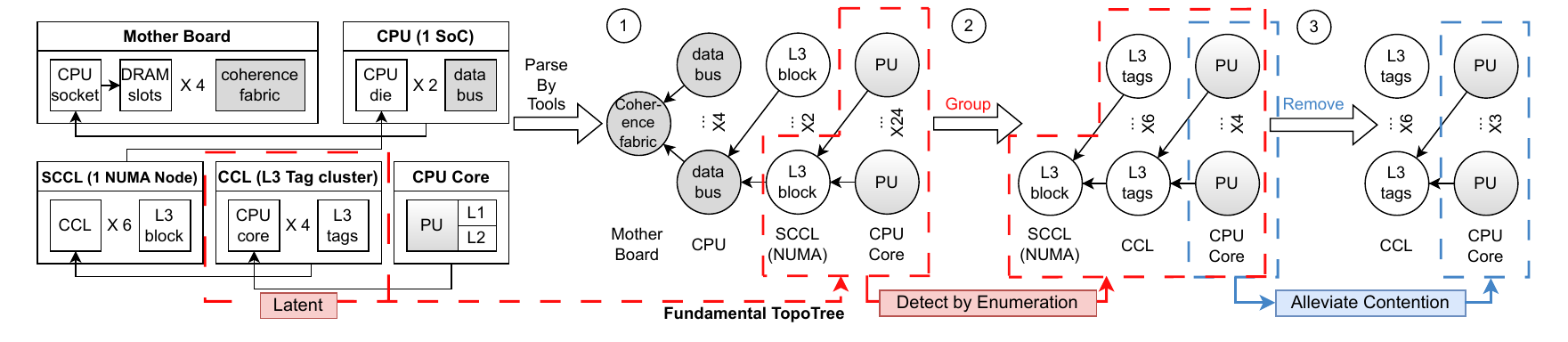}
    \vspace{-5mm}    \Description{Example TopoTree Transformation on Kunpeng920: This figure demonstrates how latent shared structures and resources are discovered and how the shared resource contention is alleviated. Tools like lscpu only reveal part of the motherboard hierarchy, such as CPU-SCCL-CPU core, while elements like the L3 cache tag are hidden from users. We treat this visible hierarchy as the foundational TopoTree. Sandwich applies group transformations to explore potential shared spatial localities, allowing the discovery of the L3 tag through a variant of the foundational tree—a process we refer to as detection by enumeration. Afterward, these trees undergo a remove transformation to alleviate potential contention along the hierarchy. This approach enables the system to manage arbitrary devices and memory hierarchies, addressing contention effectively.}
    \vspace{-4mm}
    \caption{Example TopoTree and its transformations on Kunpeng920. }
    \label{fig:kunpeng_example}
\end{figure*}

\section{Service Configuration Generation}

\subsection{TopoTree Abstraction}
\textit{TopoTree} is a multi-level tree structure 
that represents core utilization plans under a NUMA system with a shared resource hierarchy. 
As illustrated in Fig.~\ref{fig:kunpeng_example} \circled{1}, the \textit{fundamental TopoTree} is constructed based on hardware information collected through system tools like ``lstopo'' \cite{hwloc}: the leaf nodes are 
the 
processing units (PUs) where actual computation takes place, encompassing both virtual and physical CPU cores; 
non-leaf nodes 
correspond to resources shared among descendant nodes, e.g., L3 cache for Physical Units (PUs) within the same Super Core Cluster (SCCL).

\textit{TopoTree} is designed for the \textit{automation} of tree generation and optimization.
We enable efficient exploration of the latent shared structure by grouping tree nodes while mitigating resource contention through strategic node removal. 
A 
\textit{TopoTree} serves as a mapping between PUs and their associated shared resource hierarchy. It can be further utilized to represent a family of service configurations, i.e., delineating how to utilize the hardware resources including the process number, core distribution, and memory hierarchy association (e.g., NUMA information). 
Consequently, the service configuration search space is isomorphic to the domain of viable \textit{TopoTree} variants.

\subsection{Tree Transformation} 
\label{sec:transform_paradigm}

To fully explore applicable \textit{TopoTrees} (service configurations) from a \textit{fundamental TopoTree}, we design \texttt{group} and \texttt{remove} transformations on the tree. 
We have two observations regarding the service configuration and the \textit{TopoTree}:

$\triangleright$ \textbf{Symmetric and Tileable:} 
To constrain the search space, contemporary intra-op parallelism predominantly adheres to the 
SPMD paradigm~\cite{zheng2022alpa, liger, vllm}. SPMD assumes homogeneous compute capacity across cores, with instructions distributed uniformly. Such an assumption generally holds for CPUs, barring heterogeneous architectures like big.LITTLE~\cite{arm_big_little}. Consequently, this paradigm corresponds to a \textit{TopoTree} structure characterized by isomorphic subtrees. Additionally, since nodes in TopoTree are organized by clustering,
this ensures the number of PUs reachable
by any of its
direct children is equal. 


$\triangleright$ \textbf{Latent Shared Structure:} System tools like ``lstopo'' 
do not fully expose the shared structure (mainly CPU cache coherent interconnect and L3-Tags). As shown in Fig.~\ref{fig:kunpeng_example}, for a system consisting of four Kunpeng920 CPUs~\cite{taishan_v110}, \texttt{lscpu} has information that 24 cores belong to the same NUMA (L3 cache), but lacks the structure information that every CCL (4 cores) shares an L3-Tag, which might incur contention for I/O operations. 



With these observations in mind, we propose the two transformations to explore the \textit{TopoTree} (different service configurations). Given a \textit{TopoTree} with $L(d)$ nodes at level $d$, \texttt{group}$(n, t, d)$ inserts $L(d) / n$ new nodes that group $n \geq 2$ nodes each with a stride $t$ at a specified depth $d$, thereby forming a new level within the \textit{TopoTree}. \texttt{remove}$(n, d)$ eliminates $n$ right-most children of each node at the specified depth $d-1$. 
Starting from the \textit{fundamental TopoTree}, 
\sys{} first uses \texttt{group} to explore the latent shared structure to enumerate descendant TopoTrees with different possible shared structures. 
Then the \texttt{remove} transformation is applied to these explored TopoTrees (which now incorporate new node grouping relationships) to mitigate potential shared resource contention at various hierarchical levels. In Fig.~\ref{fig:kunpeng_example}, 
\circled{1} gives the \textit{fundamental TopoTree} obtained by parsing the visible hardware structure using system tools. Then in \circled{2}, some \texttt{group} transformations aggregate every four cores, forming a new hierarchical level of the TopoTree that identifies the latent L3-Tag shared structure. Subsequently in \circled{3}, a \texttt{remove} transformation eliminates specific cores, mitigating contention on the newly discovered L3-Tag structure in the explored TopoTree. 
Here \texttt{remove} can be applied 
to L3-Tag/CCL nodes to reduce L3 cache contention as well.

\sys{} systematically enumerates possible \texttt{group} and subsequent \texttt{remove} transformations until no further candidates are available. Since \texttt{group} transformations preserve parent-child relationships in the tree, their order is inconsequential.  Each descendant tree resulting from \texttt{remove} represents a new spectrum of service configurations.
Consequently, we collect intermediate \textit{TopoTrees} generated by \texttt{remove} operations, while disregarding those produced solely by \texttt{group} transformations. 
Leveraging the `Symmetric and Tileable' assumption, we apply transformations concurrently to nodes at the same hierarchical levels with the following complexity.

\noindent
\textbf{Complexity of Group Transformations.}
Let $\TG(n)$ represent the number of all possible \textit{TopoTrees} among $n$ CPU cores, $\mathtt{all\_children}(s)$ represents all underlying PUs of node $s$, and $\mathtt{cpu\_id}(c)$ represents the id of PU $c$ provided by the manufacturer. 
We make the following assumptions:
\vspace{-1mm}
\begin{assumption}\label{assum:obs_tree}
In a \textit{TopoTree}: 

\noindent\textbf{A.1}  (\emph{Symmetric}) every subgraph is symmetric;

\noindent\textbf{A.2}  (\emph{Tiled By Stride}) for an arbitrary subgraph with set $\mathbf{S}$ of 
direct children of the subgraph, there exists a stride $t \in \mathbb{N}_{>0}$ and an element $s_0 \in \mathbf{S}$ such that for any subgraph $s' \in \mathbf{S}$ and for all cores $c' \in \mathtt{all\_children}(s')$, there exists a core $c \in \mathtt{all\_children}(s_0)$ satisfying $\mathtt{cpu\_id}(c') = qt + \mathtt{cpu\_id}(c)$, where $q \in \mathbb{N}$;

\noindent\textbf{A.3} (\emph{Power of Two}): $n = 2^x$ for $x \in \mathbb{N}_{>0}$.
\vspace{-2mm}
\end{assumption}

\noindent
A.1 and A.2 hold under our observations. A.3 applies to most data center CPUs with a large number of cores.  

$\bigO(\TG(n))$ represents the upper-bound complexity of possible TopoTrees explored by \texttt{group} transformations. 
Let $\bigF(n, h)$ denote the number of all possible ways to partition $n$ subgraphs into clusters 
of size $h$. 
By our assumption and Stirling's approximation~\cite{StirlingApprox}, the following proposition outlines the complexity.

\begin{proposition}[Complexity of Group 
Transformations]

\begin{equation}\label{eq:complex_group}
\resizebox{0.85\linewidth}{!}{  
$\displaystyle
\bigO(\TG(n)) = \left\{
\begin{aligned}
&\bigO\left(\frac{n^n}{(n-1)!}\right) \approx O(e^n), &\ \text{Under A.1, A.2} \\
& \bigO\left(\frac{n^{\frac{\log_2{n}}{2}}}{(n-1)!}\right), &\ \text{Under A.1, A.2 and A.3}
\end{aligned}
\right.
$
}
\end{equation}
\vspace{-5mm}
\end{proposition}

\noindent The 
complexity is acceptable because: 1) 
the maximum number of cores in a subgraph visible by system tools is usually much smaller, $n_s \ll n$; 2) for non-datacenter CPUs, $n$ is quite small. For data center CPUs, the complexity in (\ref{eq:complex_group}) 
is limited to factorial growth in the denominator and the slower (exponential) growth in the numerator; the complexity converges to $\bigO(1)$ as $n$ increases. 
In our experimental setup, TopoTree search using \textit{group} transformations can be completed within minutes (1.18 - 600s). 

\noindent
\textbf{Complexity of Remove Transformations.}
Given a TopoTree after \texttt{group} transformations, the complexity of \texttt{remove} transformations among all possible groups of nodes 
is given by:

\begin{proposition}[Complexity of Remove 
Transformations] The complexity of \texttt{remove} on a grouped tree is $O(n^2)$.
\end{proposition} 

Detailed proofs can be found in the appendix. The complexity of \textit{remove} 
becomes vast as the number of cores increases. Fortunately, we observe opportunities to efficiently 
reduce the \texttt{remove} transformation search space: 

\noindent
\textit{
1) Equivilant TopoTrees:} Different \texttt{remove}
transformation paths can lead to common intermediate tree representations. Inspired by the Merkle tree used in blockchain~\cite{merkle}, we generate a header hash for each tree based on its structure and node features. A \textit{TopoTree} 
is only considered for a subsequent \texttt{remove} transformation if it is new.

\noindent
\textit{2) Transformation Tree:} While \texttt{remove} transformations initially enhance performance by mitigating resource contention, they may eventually degrade efficiency due to reduced computation parallelism. If a descendant tree from a \texttt{remove} transformation yields no performance improvement over its parents, its further descendants are pruned. Successive \texttt{remove} transformations on a parent tree constitute a transformation tree, as illustrated in Fig.~\ref{fig:wf_sandwich} ("tree of transformation"). At depth $d$ of the transformation tree, \texttt{remove} candidates are determined by the number of nodes at $d-1$. We employ topological sorting to traverse
the transformation tree from the root. If any node is discardable,
all its descendants are eliminated.

\subsection{
TopoTree to Service Configuration}\label{sec:interpret}

\begin{figure}
    \centering
    \includegraphics[width=0.75\linewidth]{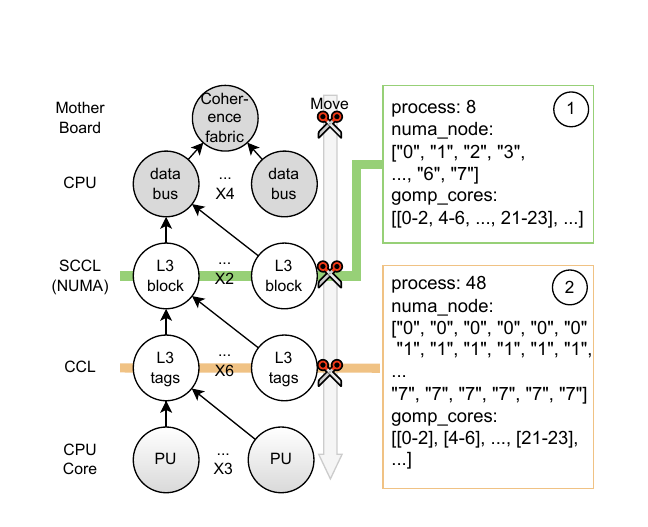}
    \Description{Example TopoTree Interpretation. Consider a TopoTree structure represented as Motherboard x1-CPU x4-SCCL(NUMA) x2-CCL x6-CPU core x3. The interpretation of this structure depends on the chosen cross-section level. (1) When the cross-section is at the SCCL level, it generates a configuration with 8 processes, NUMA node assignments [0,1,2,3,4,5,6,7], and core clusters consisting of 3 out of 4 cores in each NUMA node. (2) Alternatively, if the cross-section is at the CCL level, the configuration results in 48 processes, with NUMA node assignments [$8\times$ '0', $8\times$ '1', $\ldots$, $8\times$ '7'], and each process allocated 3 out of 4 CPU cores. }
    \caption{Example TopoTree Interpretation}
    \label{fig:example-interpret}
    \vspace{-5mm}
\end{figure}

CPU cores can be grouped or partitioned at either the \textbf{thread} or the \textbf{process} level. As a result, different model parallelization strategies can be executed on the same \textit{TopoTree}. As illustrated in Fig.~\ref{fig:example-interpret}, 144 cores can be configured in two ways: \circled{1} 8 processes (corresponding to TP=8 in our scenario) with 16 cores each, or \circled{2} 48 processes with 3 cores each

To enumerate candidate service configurations from a TopoTree, we propose the \textbf{sandwich-config} algorithm. This algorithm 
starts at the tree's root and progresses level-wise, applying a horizontal cross-section (as depicted in Fig.~\ref{fig:example-interpret}). Nodes intersecting this cross-section determine the process count and their associated memory hierarchy (e.g., NUMA), while the subtrees beneath these nodes delineate core spatial locality and utilization strategies. 
The complexity of the \textbf{sandwich-config} algorithm is constrained by the TopoTree height, which in turn depends on the number of cores. With a maximum height of 
$\log_2(k)$, where $k$ is the number of cores,
the algorithm's complexity is $O(\log_2(k)$. In practice, this value is typically less than 10. We further propose several practical heuristics to speed up service configuration generation. 

\noindent
\textit{
1) Equivalent Configuration:} Hashing and eliminating the same configurations met during the algorithm procedure. 

\noindent
\textit{
2) Operator Limitations:} Certain operators restrict the possible tensor-parallel degrees. 
For example, the TP degree of attention must be divisible by the number of KV-heads and query heads, 
which puts a limit to the largest TP. 

\noindent
\textit{
3) Early Stopping:} 
We further enable early stopping, using a patience score to halt when consecutive non-optimal results are reached. 
We group service configurations with the same process number and NUMA information, and introduce early stopping within each group. If early stopped, all configurations in that process group are discarded.

We determine the optimal service configurations in terms of end-to-end latency
through latency simulation based on sampled token generation traces. 
It is noteworthy that kernel performance can significantly impact service configuration outcomes due to its direct influence on latencies.
However, simultaneously identifying the best service configuration and tensor schedule would lead to very high search complexity. For efficiency, when identifying the best service configurations, we adopt a default tensor schedule generated by a cost model-based polymerization method~\cite{dietcode}. 
Then we provide top-k best service configurations among all TopoTrees for subsequent kernel orchestration. In our experiments, this approach has demonstrated both performance superiority 
and efficiency with $k=10$.

\begin{figure*}[t]
    \centering
    \includegraphics[width=\textwidth]{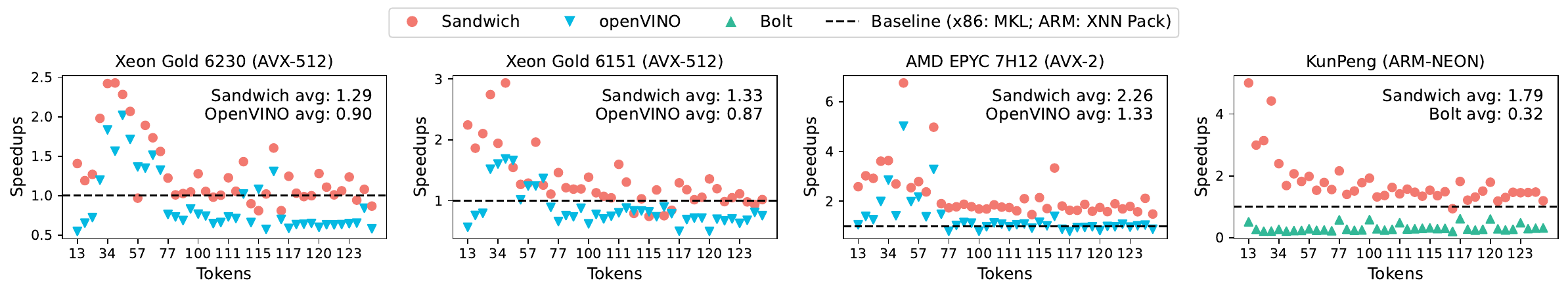}
    \vspace{-6mm}
    \Description{Kernel execution comparison between \sys{} and other vendor solutions. We select MKL as the x86 baseline and XNN pack as the ARM baseline to evaluate speedup. Performance improvements are assessed on four different devices: Xeon Gold 6230, Xeon Gold 6151, AMD EPYC 7H12, and Kunpeng. The x-axis represents a series of GEMM operators (4 for each token size), y-axis is the speed up. Graph illustrates normalized kernel performance. On Xeon Gold 6230, \sys{} achieves an average 1.29$\times$ speedup, while OpenVINO shows 0.9$\times$. For Xeon Gold 6151, \sys{} attains a 1.33$\times$ speedup, with OpenVINO at 0.87$\times$. On AMD EPYC 7H12, \sys{} demonstrates a 2.26$\times$ speedup, compared to OpenVINO's 1.33$\times$. Finally, on Kunpeng, \sys{} exhibits a 1.79$\times$ speedup, while Bolt achieves 0.32$\times$.}
    \vspace{-2mm}
    \caption{Comparison of kernel execution speed-up among \sys{} and vendor solutions.}
    \label{fig:kernel-vendor}
    \vspace{-1mm}
\end{figure*}

\section{Tensor Program Generation}


Given a 
model partition plan, the prefill shapes are known except for the input sequence length. 
This allows us to generate tensor programs for all possible shapes in the inference process, considering input sequence length up to the maximum allowed by the model. 
We design a {\bf sandwich-kernel} approach to generate tensor schedules for dynamic shapes under a specific service configuration.
Besides using the fast-start-and-finetune strategy, \sys{} also utilizes the similarity of the LLM prefill shapes to advocate a sliding window technique and tensor program reuse to reduce tuning time. 


\noindent
\textbf{Micro-Kernel Generation.} \sys{} generates MKs of size $\mu_M \times \mu_N$, where $\mu_M$ and $\mu_N$ rows of data along the reduction dimension $K$
are loaded, multiplied, and accumulated using SIMD registers. 
Similar to Roller~\cite{Roller}, \sys{} generates MKs tailored to align with CPU processing units, 
particularly focusing on avoiding register spilling, which can degrade performance.
As modern CPUs are typically equipped with 32 vector registers, \sys{} enumerates potential MK candidates utilizing up to 32 registers. The size of reduction dimension $b_K$ is also aligned with the cacheline size, which is the maximum number of bytes loaded into a cache simultaneously.


\noindent
\textbf{Fast-start and 
finetune.}
Similar to 
\cite{Roller, dietcode, milkpoly}, \sys{} constructs a tensor program by scaling up computation slices along the cache hierarchy with every MK from the previous step. To efficiently utilize the initial steps and simultaneously search for concurrent thread polymerization and replicable computation slices, we propose a two-phase approach: a fast initial search followed by fine-tuning.

1) \textit{Fast Start:} 
Like TCP~\cite{tcp_control}, we expand the shape of a computation slice 
with an exponentially growing step of $2^{t_D} \times \text{tilesize}$, where $t_D$ is the accumulated expansion step size on dimension $D$. The performance 
(GFLOPS) of the growing kernel is collected by profiling at each step.
If any expansion on a dimension $D$ fails to provide performance gains or degrades performance, the size of the computation slice is rolled back to the previous step along this dimension. To avoid reducing the search space for 
polymerization schemes, 
we stop expanding a dimension if doing so limits its parallelizability. 
For example, for $(M,N,K)=(128, 64, 64)$ and $\text{nthread}=2$, 
the maximum size of the computational slice 
should be limited to $(64, 32, 32)$. 

2) \textit{
Finetune:}
After fast start, we enumerate thread polymerization schemes utilizing all CPU cores available to the tensor program. Under each scheme, we grow the computational slice
along the dimension that provides the most efficiency, until the best computation slice is found. Finally, we compare the efficiency of the result tensor programs under every polymerization scheme and MK to choose the best one.

\noindent
\textbf{Micro-kernel sliding window \& tensor schedule reuse.}
Since many input shapes in the prefill workload shares the same sizes for dimensions except the one representing sequence length, 
we process such input shapes in groups and order them by the size of the input dimension, starting from 1. We propose two optimizations to speed up the tuning: 

1) {\em Sliding window:}
For a large enough sequence length, the MK yielding the optimal tensor program is always the one that exploits data reuse the most with vector registers, but when the input sequence length is relatively small, the input shape is skewed and requires a diverse set of candidate MKs. We create a sliding window of size $\sigma$. For the first $\sigma$ shapes, we use all MKs as candidates; for the following shapes, we only consider the $\sigma$-last chosen MKs. 
This 
helps us identify and prioritize the most effective MKs for larger token sizes while sustaining a diverse selection of MKs for small token sizes. In our experiments, a $\sigma$ of 16 yields a good balance between scheduling overhead and tensor schedule quality. 


2) {\em Tensor Schedule Reuse:} 
As the token size continues to grow, the polymerization scheme and the shape of the computation slice tend to stabilize. This stability occurs because the shape becomes less skinny, and
large enough to obtain optimal computation slices without reducing the search space of polymerization. 
We set an empirical early-stopping criterion for tensor schedule searches. When the new optimal slice's difference is within a threshold of the previous average, the slice will be fixed for the following shapes. The same policy applies to threads. In addition, we cache the searched schedules 
to avoid repetition across different shape groups. With the optimal slice and polymerization, \sys{} can easily generate a schedule for the unseen larger shape by dividing the input shape with the polymerization plan and repeating the computation slice on each division. 

\begin{figure}[t]
    \centering
    \includegraphics[width=\linewidth]{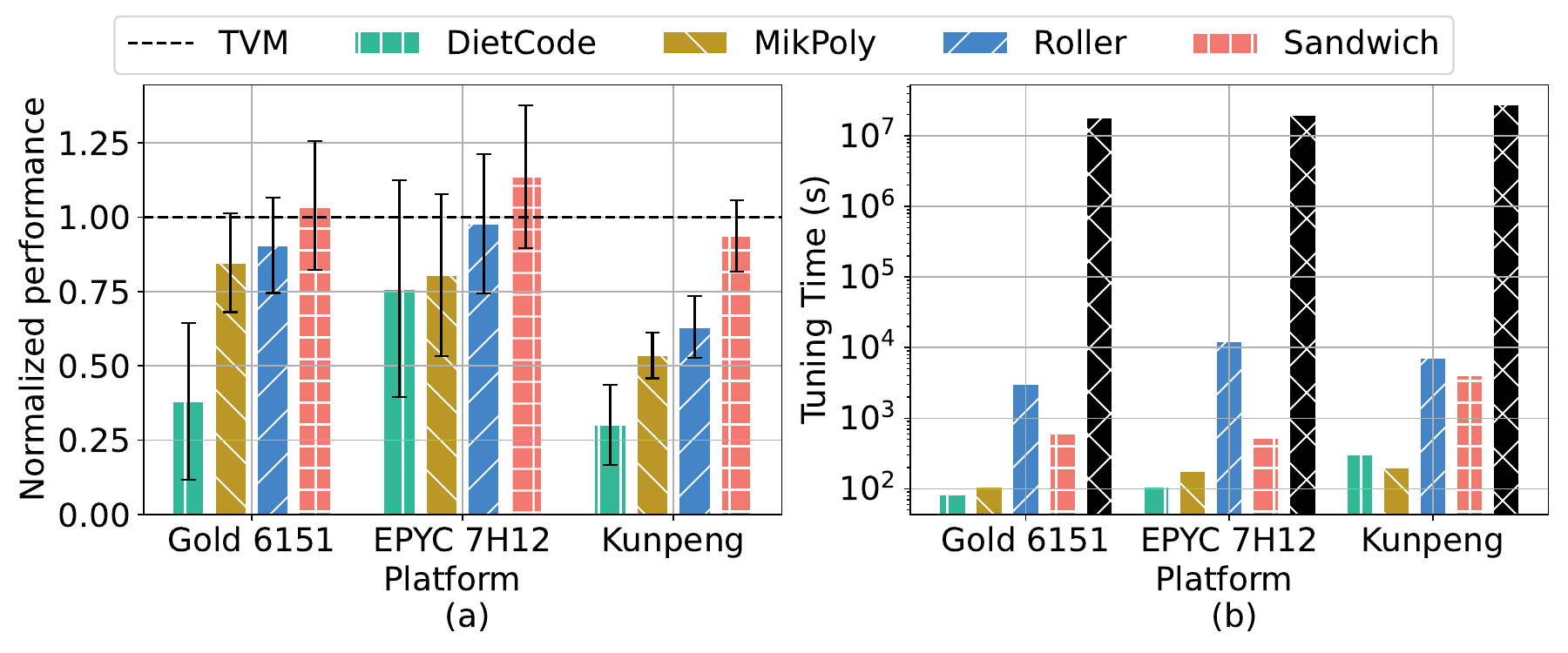}
    \vspace{-3mm}
    \Description{Kernel generation comparison between \sys{} and other compilers. In graph (a), the x-axis represents the test platforms: Xeon Gold 6151, AMD EPYC 7H12, and Kunpeng. The y-axis shows the normalized performance relative to TVM. Results demonstrate that \sys{} outperforms other solutions in all cases, surpassing TVM in the first two scenarios and achieving comparable performance in the third. Graph (b) compares the tuning time of all solutions, with the y-axis representing tuning duration. We observe that \sys{} is three orders of magnitude faster than TVM in tuning. While slower than cost model-based solutions, it remains faster than scale-up and scale-out approaches.}
    \vspace{-5mm}
    \caption{Kernel generation comparison among \sys{} and other compilers: (a) Normalized kernel performance; (b) Tuning time.}
    \vspace{-2mm}
    \label{fig:kernel-compiler}
\end{figure}

\begin{figure*}[t]
    \centering
    \begin{subfigure}{0.47\linewidth}
        \centering
        \includegraphics[width=\linewidth]{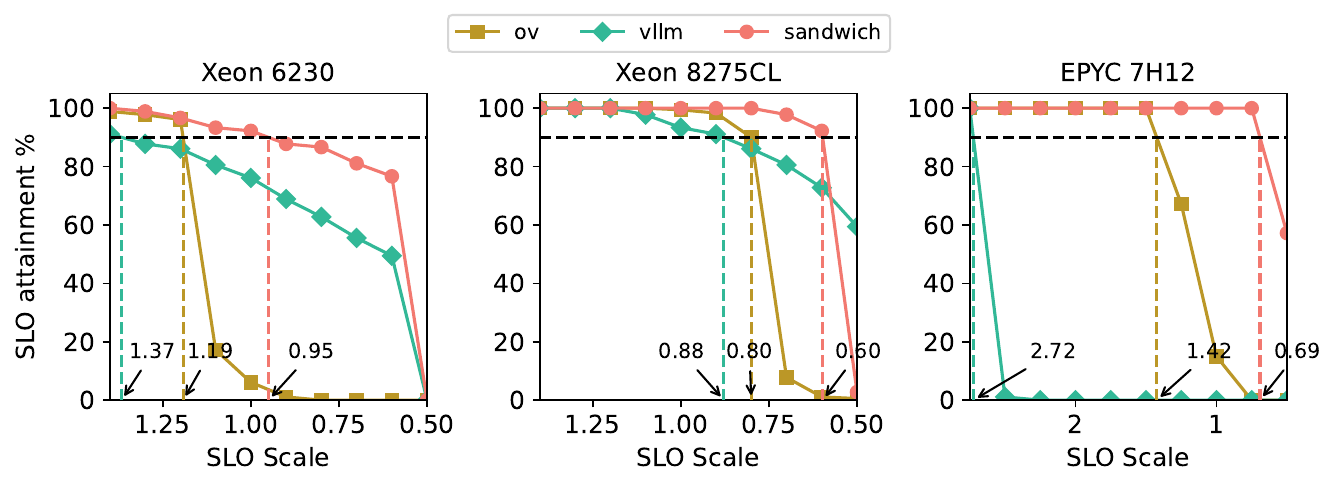}
        \caption{Llama-1.3b, ShareGPT}
        \label{fig:1.3b}
    \end{subfigure}
    \hfill
    \begin{subfigure}{0.47\linewidth}
        \centering
        \includegraphics[width=\linewidth]{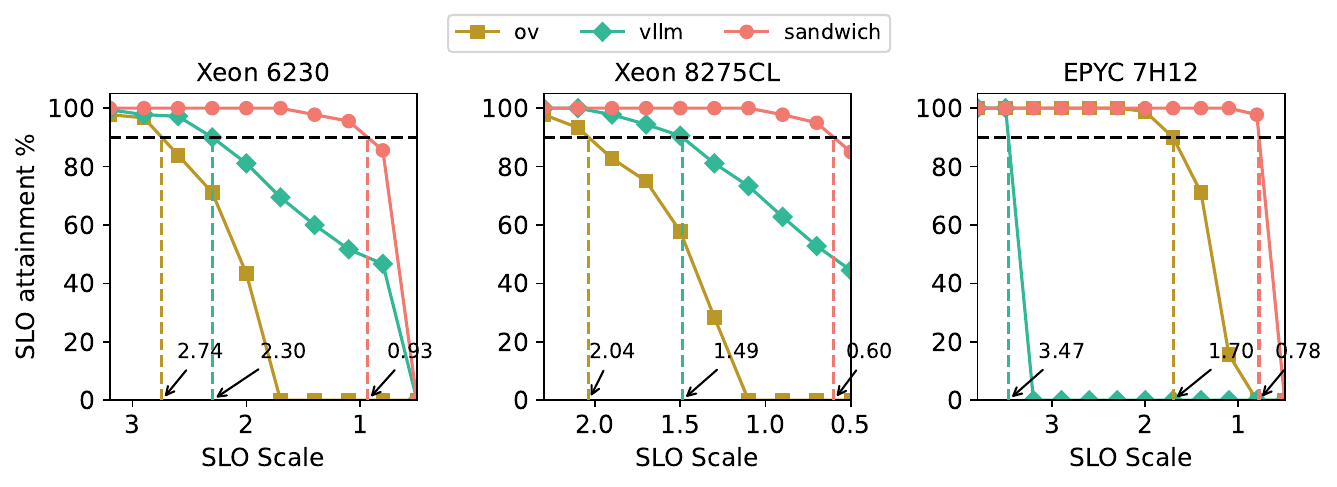}
        \caption{Llama3-8b, ShareGPT}
        \label{fig:8b}
    \end{subfigure}
    \vspace{-3mm}
    \caption{Comparison of SLO attainment percentage under different SLO scales.}
    \Description{Six figures in a two-by-three grid are comparing SLO retainment percentage and output token throughput of two different llama models on two input datasets over five different CPU systems. Subfigures a and b contain line plots of SLO attainment percentages of llama 1.3b and llama 3 8b, with inputs selected from the sharegpt datasets. Subfigures a and b each contain three sub-subfigures, showing results form Xeon 6230, Xeon 8275CL, and EPYC 7H12. Each sub-subfigure contains three lines, showing how the SLO attainment rate as the SLO scale decreases. All of the lines start out flat and curve down until it reaches zero. The sandwich line hits the 90 percent horizontal line farthest to the right, maintaining the most stringent SLO, while ov performs better than vllm in the 1.3b model, but worse in the 8b model. Subfigures c to f are put into a two-by-two grid, covering output token per second for two llama models of size 1.3b and 8b on inputs from two datasets, share-gpt and lmsys-chat. Each subfigure contains four groups of bars that represents different backends on different CPU hardware, described in section 8. Sandwich outperforms other solutions on every platform and every model. Although performing worse than sandwich, Vllm is competitive with small models; ov is competitive with larger models. }
    \label{fig:main_exp}
\end{figure*}

\section{Dual-IR and Runtime Implementation}


\sys{} is implemented with 14,400 lines of Python code and 6,700 lines of C++ code. It is fully integrated with vLLM~\cite{vllm}, incorporating optimizations such as continuous batching~\cite{orca}, and can be directly invoked using vLLM commands. Kernel optimizations like positional embedding~\cite{rope}, FlashAttention~\cite{dao2022flashattention}, and PagedAttention~\cite{vllm} have been parsed to IR and incorporated into our codebase, ensuring seamless integration and support across multiple devices.  


\subsection{Dual-IR}\label{sec:dual-ir}
Specialized hand-crafted kernels, such as attention kernels, typically necessitate labor-intensive reimplementation on a target hardware platform and compiler DSL. 
To expose and tune customized key parameters while ensuring portability from existing tensor programs, \sys{} kernel compiler 
adopts a dual-level IR architecture: a low-level IR with a syntax resembling the low-level programming language used for LLM kernels (e.g., C++) for easy adoption, 
and another high-level IR with Python-like syntax for abstracting SIMD instructions and easier insertion of key parameters (in the case of our GEMM tensor schedule, encompassing $\mu_M, \mu_K$ in MKs). 
Complex hand-crafted tensor programs are first parsed to our low-level IR,
and then to the Python-like IR for tuning. 

\subsection{Rank-Shifted Share Memory Communication}


We implement efficient inter-process communication using shared memory (SHM). The master process obtains a file descriptor and distributes it to worker processes, which then cache and map it for shared data access. During all-reduce operations, workers wait for the master to complete data copying, accumulate their data in shared memory, and copy the result back to their local memory. Despite  its efficiency for data sharing \cite{xFasterTransformer}, this approach faces challenges in handling concurrent accesses and variable data sizes.

We propose a rank-shifted adaptive SHM communication approach, inspired by the geometric circular shift method from~\cite{rajasekaran2023cassini}. This approach assigns a unique rank-based shift to each process, creating a circular writing pattern that enables concurrent execution without waiting. We dynamically adapt the shared memory block size, ensuring it exceeds the cache-line size to avoid false sharing~\cite{hoard}.

\begin{figure*}[t]
    \centering
    \begin{subfigure}{0.47\linewidth}
        \centering
        \includegraphics[width=\linewidth]{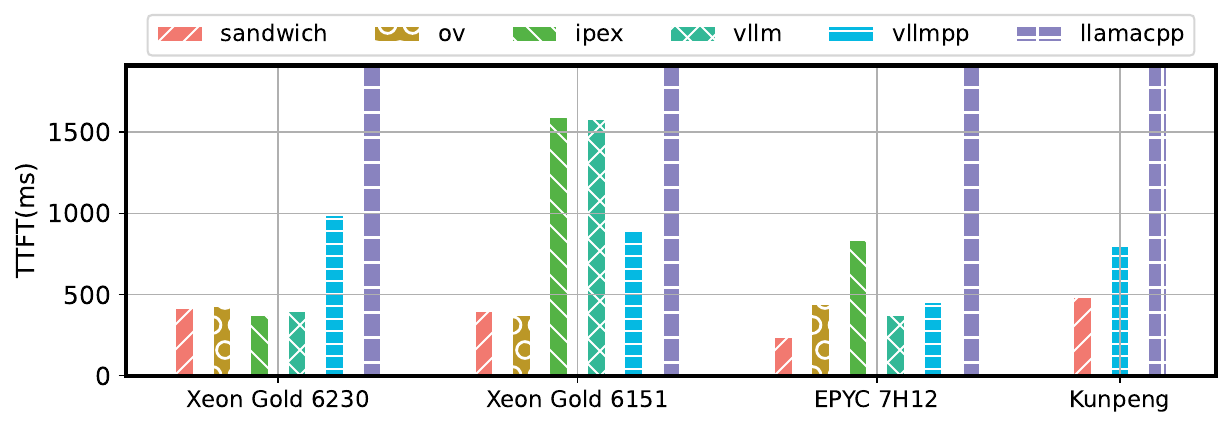}
        \caption{Llama-1.3b, ShareGPT}
    \end{subfigure}
    \hfill
    \begin{subfigure}{0.47\linewidth}
        \centering
        \includegraphics[width=\linewidth]{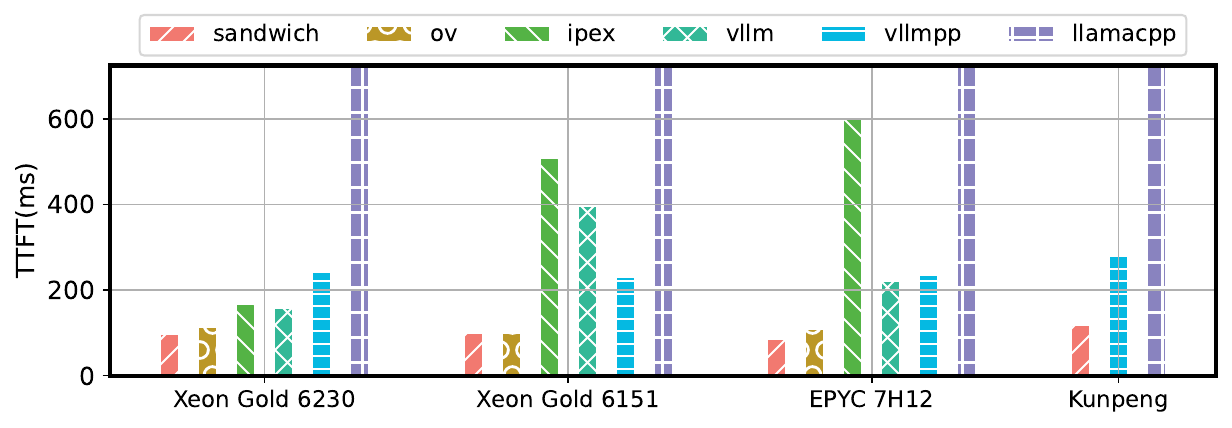}
        \caption{Llama-1.3b, LMsys-Chat}
    \end{subfigure}
    
    \vspace{2mm}
    
    \begin{subfigure}{0.47\linewidth}
        \centering
        \includegraphics[width=\linewidth]{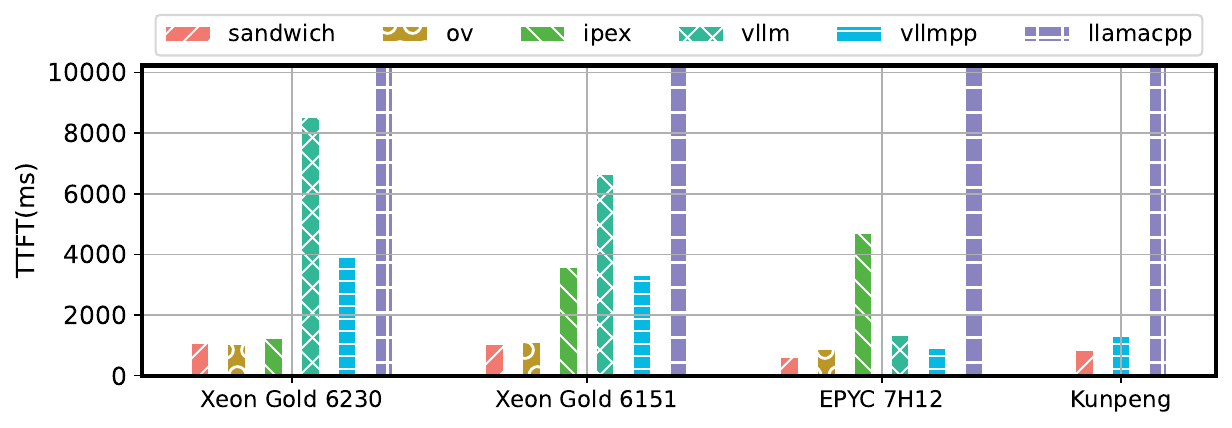}
        \caption{Llama3-8b, ShareGPT}
    \end{subfigure}
    \hfill
    \begin{subfigure}{0.47\linewidth}
        \centering
        \includegraphics[width=\linewidth]{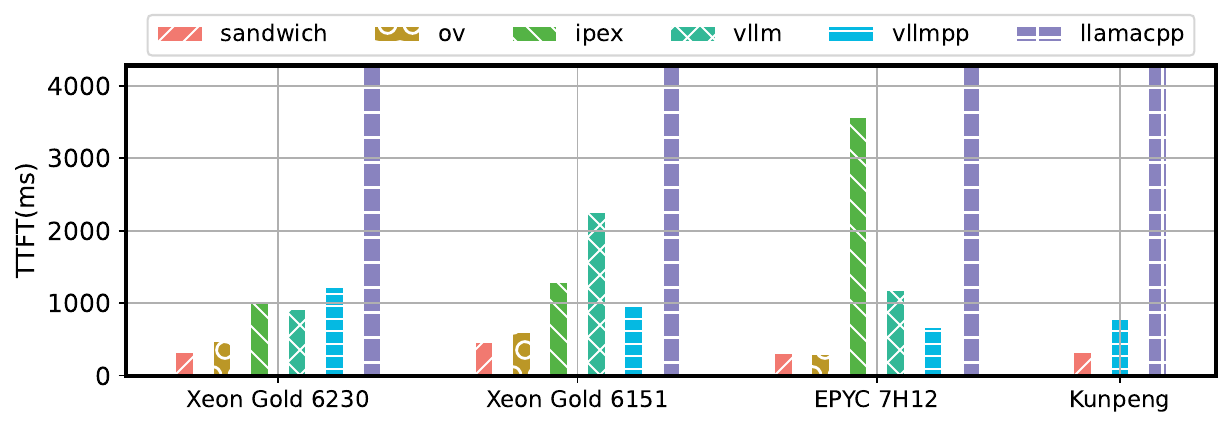}
        \caption{Llama3-8b, LMsys-Chat}
    \end{subfigure}
    \vspace{-2mm}
    \caption{Comparison of TTFT in single sequence serving.}
    \Description{Six figures in a two-by-three grid are comparing SLO retainment percentage and output token throughput of two different llama models on two input datasets over five different CPU systems. Subfigures a and b contain line plots of SLO attainment percentages of llama 1.3b and llama 3 8b, with inputs selected from the sharegpt datasets. Subfigures a and b each contain three sub-subfigures, showing results form Xeon 6230, Xeon 8275CL, and EPYC 7H12. Each sub-subfigure contains three lines, showing how the SLO attainment rate as the SLO scale decreases. All of the lines start out flat and curve down until it reaches zero. The sandwich line hits the 90 percent horizontal line farthest to the right, maintaining the most stringent SLO, while ov performs better than vllm in the 1.3b model, but worse in the 8b model. Subfigures c to f are put into a two-by-two grid, covering output token per second for two llama models of size 1.3b and 8b on inputs from two datasets, share-gpt and lmsys-chat. Each subfigure contains four groups of bars that represents different backends on different CPU hardware, described in section 8. Sandwich outperforms other solutions on every platform and every model. Although performing worse than sandwich, Vllm is competitive with small models; ov is competitive with larger models. }
    \label{fig:main_exp_ttft}
\end{figure*}

\section{Evaluation}

\subsection{Experimental Settings}
\noindent 
\textbf{Platform.}
We evaluate \sys{} across various NUMA systems. For x86 architectures, 
we use (i) $2\times$ Intel(R) Xeon(R) Gold 6151 CPU (2 NUMA nodes) with 18 physical cores (PC) / 36 virtual cores (VC) each~\cite{wiki:Intel_Xeon_Skylake}, (ii) $2\times$ Xeon Gold 6230 CPU with 20 PC/ 40 VC each (2 NUMA nodes)~\cite{intel_xeon_gold_6230}, (iii) $2\times$ Xeon Platinum 8272CL CPU with 24 PC/ 48 VC each (2 NUMA nodes)~\cite{8272CL},
and (iv) $2\times$ AMD EPYC 7H12 with 64 PC / 128 VC each (2 NUMA nodes). The prior two CPUs have AVX-512 vector extension. The latter one has AVX-2 vector extension. 
For aarch64 (ARM), we assess 
$4\times$ Kunpeng 920 CPUs with 8 NUMA nodes, 192 PCs, and NEON instructions\cite{kunpeng920}. 

\noindent
\textbf{Model Precision.} We employ BF16 for experiments 
on AVX-512 CPUs, and FP32 for those on AVX-2 and Kunpeng platforms. Notably, since TVM does not support BF16 tuning on CPUs, we utilize FP32 for kernel experiments 
for a clear comparison across different CPU platforms and compilers.

\noindent 
\textbf{Model and workload.}
We run the Llama series~\cite{Touvron2023LLaMAOA}, a prominent family of LLMs. 
For dynamic-shape benchmarks, we create a payload generator to produce all possible shapes of the operator for a model, given the maximum sequence length
and model configuration, such as hidden size, intermediate size, KV, and query head numbers. The input sequence lengths
are sampled within a specified range to generate the test payloads, and we explicitly use cores on a single NUMA node for these experiments to avoid interference. We tune all possible shapes with maximal sequence length $seq_{max}$ for compiler comparison.  

For serving, we evaluate both single sequence and batched serving. We generate serving workloads based on the chatbot dataset ShareGPT and LMsys-Chat-1M~\cite{lmsys}. 
For single sequence serving, we sample 90 test requests from the datasets 
and feed them into the serving system sequentially. For batched serving, request arrival times are generated using a Poisson distribution with varying request rates 
\cite{zhong2024distserve, vllm}. 

\noindent
\textbf{Metrics.}
For dynamic-shape benchmarks, we measure operator latency speed-ups (with performance normalized to that of TVM) and tuning time.
For serving, 
we adopt 
SLO attainment as our primary 
metric. Additionally, 
we measure (1) the average token generation throughput during single-sequence serving, and (2) Goodput for batched serving, defined as the maximum request rate that can be sustained while meeting the SLO attainment goal. 
Consistent with the previous study~\cite{zhong2024distserve}, we set the SLO attainment goal to 90\% (i.e., P90). 
The SLO parameters are empirically determined as described in Sec.~\ref{sec:serve}.  

\noindent
\textbf{Baselines.} 
The dynamic-shape solutions that we compare with are: (1) standard GEMM operators from DNN frameworks and vendor-specific solutions, including MKL-DNN~\cite{intelmkl} and OpenVINO~\cite{openvino} for x86, and Bolt~\cite{bolt} and Torch with XNNPACK~\cite{xnnpack} on ARM SoC; (2) compiler solutions such as TVM Ansor~\cite{Ansor}, and dynamic-shape solutions DietCode, Roller, and MikPoly~\cite{dietcode, Roller, milkpoly}. Since these solutions are originally designed for GPUs, we reimplement them for CPUs by translating their scheduling policies. We provide them the same MK candidates used by \sys{}, and only differ from them in polymerization. 
AutoTVM~\cite{autoTVM} 
does not provide an official code 
for CPU linear ops, thus not included.

We also compare \sys{}  with various serving solutions. 


\noindent
$\bullet$ Hardware-specific vendor solutions: (1) intel-extensiton-for-pytorch (ipex)~\cite{IntelCPUInference}, and (2) OpenVINO (ov)~\cite{openvino}.\\ 
 $\bullet$ GPU-serving solutions: (3) vLLM~\cite{vllm} with its default CPU backend and no model partition; (4) vLLM++ (vllmpp)~\cite{zhong2024distserve}, a vLLM version that enumerates TP strategies for optimal performance, where NUMA node and cores are partitioned evenly (to distinguish it from vLLM, we set vllmpp with a minimum TP size of 2). \\
 $\bullet$ SOTA open-source solution (5) llama.cpp~\cite{llama.cpp}. 
 
 For \sys{}, we obtain $topk=10$ service configurations and identify the best performance among them. 
We do not compare with xFasterTransformer~\cite{xFasterTransformer} as it is mainly optimized for advanced instructions with at least AVX512-bf16, and offers poor support to our CPU platforms.


\begin{figure*}[t]
    \centering
    \begin{subfigure}{0.47\linewidth}
        \centering
        \includegraphics[width=\linewidth]{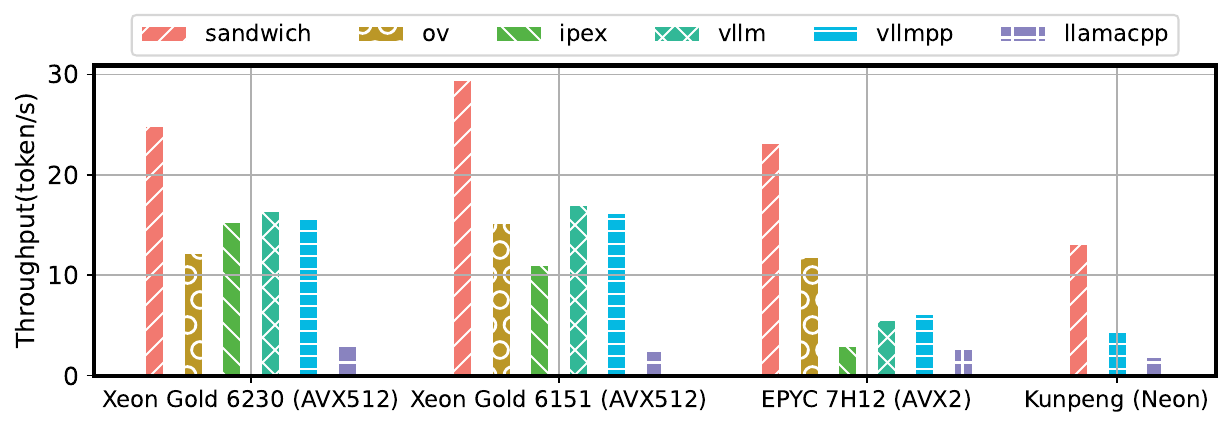}
        \caption{Llama-1.3b, ShareGPT}
    \end{subfigure}
    \hfill
    \begin{subfigure}{0.47\linewidth}
        \centering
        \includegraphics[width=\linewidth]{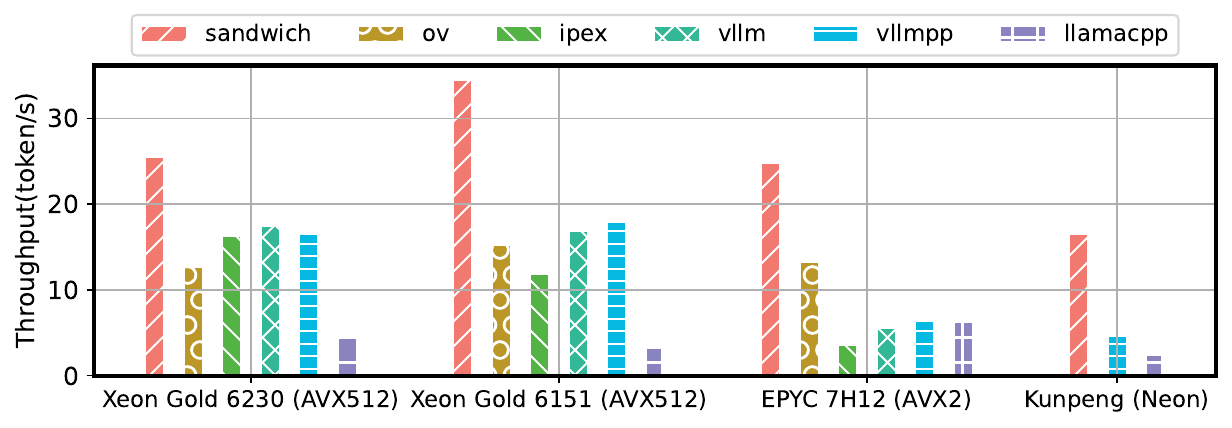}
        \caption{Llama-1.3b, LMsys-Chat}
    \end{subfigure}

    \vspace{2mm}

    \begin{subfigure}{0.47\linewidth}
        \centering
        \includegraphics[width=\linewidth]{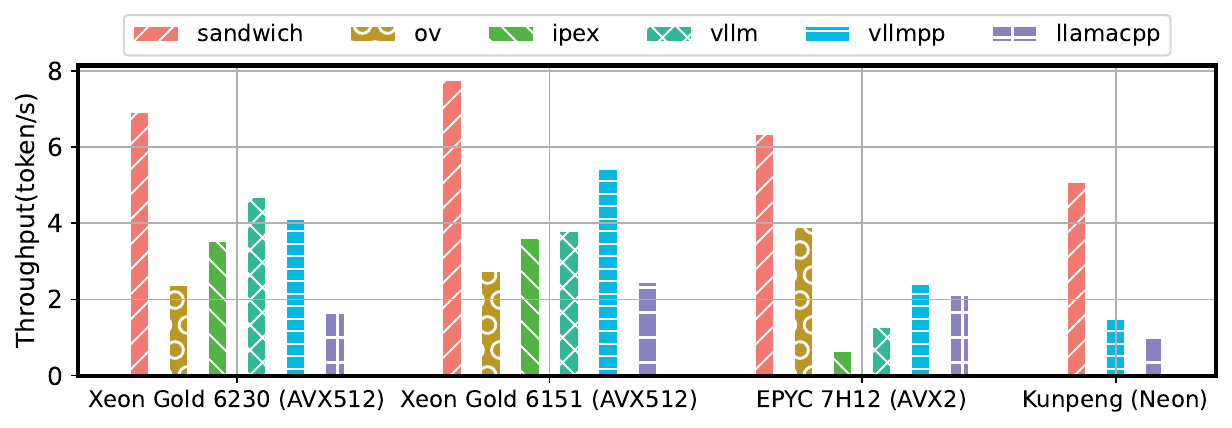}
        \caption{Llama3-8b, ShareGPT}
    \end{subfigure}
    \hfill
    \begin{subfigure}{0.47\linewidth}
        \centering
        \includegraphics[width=\linewidth]{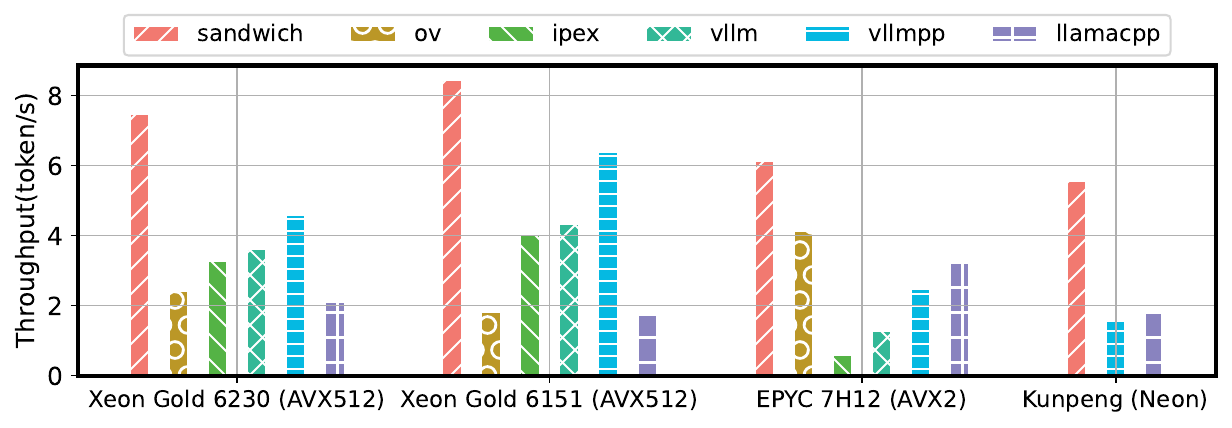}
        \caption{Llama3-8b, LMsys-Chat}
    \end{subfigure}

    \vspace{-2mm}
    \caption{Comparison of output token generation throughput for single sequence serving.}
    \Description{Six figures in a two-by-three grid are comparing SLO retainment percentage and output token throughput of two different llama models on two input datasets over five different CPU systems. Subfigures a and b contain line plots of SLO attainment percentages of llama 1.3b and llama 3 8b, with inputs selected from the sharegpt datasets. Subfigures a and b each contain three sub-subfigures, showing results form Xeon 6230, Xeon 8275CL, and EPYC 7H12. Each sub-subfigure contains three lines, showing how the SLO attainment rate as the SLO scale decreases. All of the lines start out flat and curve down until it reaches zero. The sandwich line hits the 90 percent horizontal line farthest to the right, maintaining the most stringent SLO, while ov performs better than vllm in the 1.3b model, but worse in the 8b model. Subfigures c to f are put into a two-by-two grid, covering output token per second for two llama models of size 1.3b and 8b on inputs from two datasets, share-gpt and lmsys-chat. Each subfigure contains four groups of bars that represents different backends on different CPU hardware, described in section 8. Sandwich outperforms other solutions on every platform and every model. Although performing worse than sandwich, Vllm is competitive with small models; ov is competitive with larger models. }
    \label{fig:main_exp_throu}
\end{figure*}

\subsection{Dynamic-Shape Benchmark}
In this section, we compare the performance of \sys{}-generated kernels against baseline methods on commonly used shapes. Specifically, we conduct a case study on GEMM, which is the primary computational operator in LLMs.

\noindent
\textbf{Vendors.}
We first compare with 
vendor solutions for x86 and ARM architectures, namely MKL and XNNPack. Fig.~\ref{fig:kernel-vendor} illustrates the speedups of the \sys{} kernel relative to the vendor and third-party libraries. The x-axis specifies the token size, each representing a set of all possible GEMM operations conducted during serving (e.g., QKV, FFN) in a Llama-1.3b model. 
For each kernel, we perform a warmup of 5 times and repeat the measurement 100 times to obtain average results. 

\sys{} achieves  
$1.27-4.02\times$ speedup compared to the baselines, and outperforms other solutions (e.g., OpenVINO and Bolt) in most cases. Notably, \sys{} demonstrates more efficiency for kernels with smaller M-size $(M < 64)$; when the token size is larger (token size $> 100$), the speedup over different solutions tends to decrease. We attribute this to the fact that as M increases, the kernel becomes less skinny, allowing vendor solutions to perform well.

\noindent
\textbf{Compilers.}
We further compare 
kernel program performance normalized against TVM 
and tuning time over Gold 6151, EPYC 7H12, and Kunpeng platforms. For AutoTVM~\cite{autoTVM}, we set the number of tuning trials 
to 800, as recommended in their official documentation, and allow early stopping.
Due to its extended tuning time, we sample only 20 shapes to tune with TVM. We range the token size in $[1, seq_{max}]$, where $seq_{max}$ is specified in the model configuration 
in the tuning process. Fig.~\ref{fig:kernel-compiler}(a) shows that \sys{}'s kernel achieves better or comparable performance relative to the TVM tuner, while Fig.~\ref{fig:kernel-compiler}(b) reveals that \sys{} requires less than 90\% of the tuning time, as compared to TVM. \sys{} is faster than Roller in terms of tuning time due to its quick start and effective tensor-schedule reuse. Cost-model-based methods, such as DietCode and MikPoly, are extremely fast in tuning but fail to provide competitive kernel performance. \sys{} outperforms all other compilers because it jointly optimizes
both concurrent polymerization and parallelizable schemes.  

\begin{figure}[t]
    \centering
    \begin{subfigure}{0.45\linewidth}
        \centering
        \includegraphics[width=\linewidth]{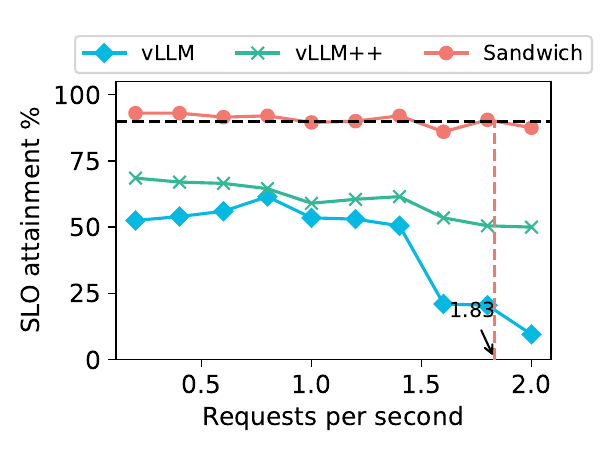}
        \caption{Llama-160m, Xeon Gold 6151}
    \end{subfigure}
    \hfill
    \begin{subfigure}{0.45\linewidth}
        \centering
        \includegraphics[width=\linewidth]{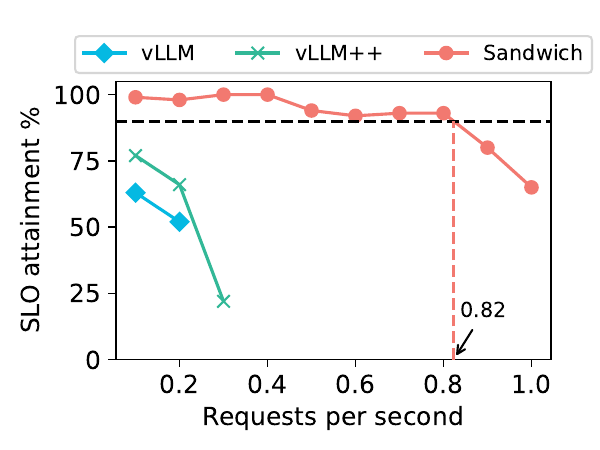}
        \caption{Llama-1.3b, Xeon Gold 6230}
    \end{subfigure}

    \vspace{2mm}

    \begin{subfigure}{0.45\linewidth}
        \centering
        \includegraphics[width=\linewidth]{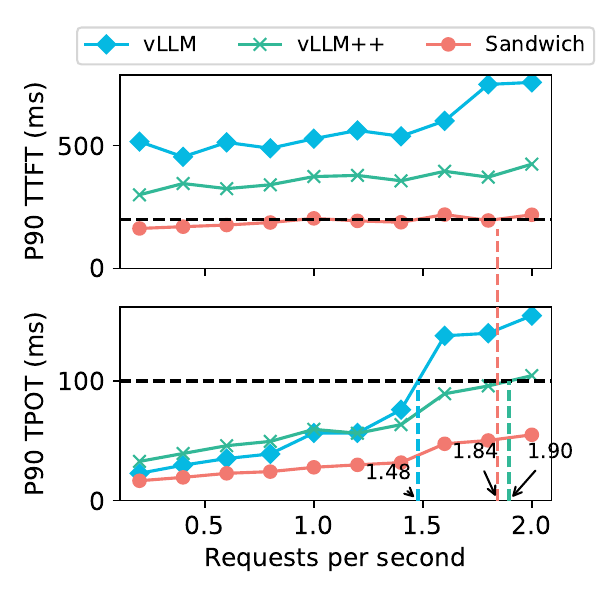}
        \caption{Llama-160m, Xeon Gold 6151}
    \end{subfigure}
    \hfill
    \begin{subfigure}{0.45\linewidth}
        \centering
        \includegraphics[width=\linewidth]{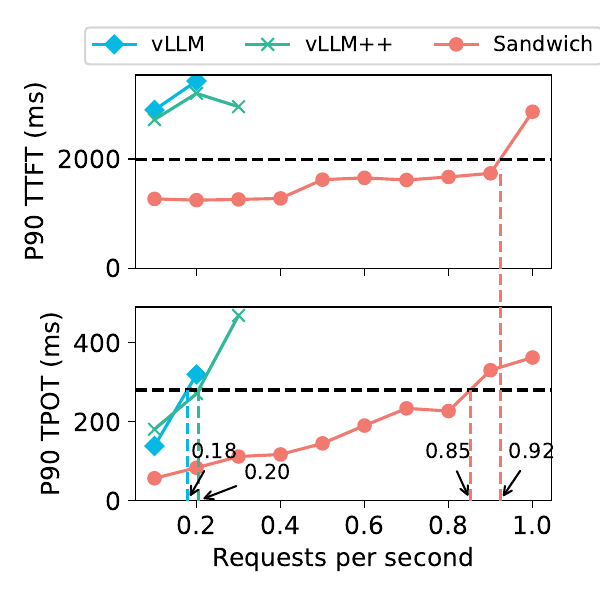}
        \caption{Llama-1.3b, Xeon Gold 6230}
    \end{subfigure}

    \vspace{-4mm}
    \caption{SLO and Goodput comparison for batched serving.}
    \Description{A figure containing four sub-figures showing SLO and goodput comparison for batched serving. Sub-figure a and b contain line plots with the change in SLO attainment percentage for the llama-160m model on Xeon Gold 6151 and the llama 1.3b model on Xeon Gold 6230, respectively. There are three lines representing vLLM, vLLM++, and sandwich. Sandwich is able to retain 90 percent SLO with a request rate of 1.83 requests per second (rps) for the 160m model, and 0.82 rps for the 1.3b model. The other two backends fail to meet SLO for all the request rates tested. All lines start flat and curve down. Sub-figure c and d contain line plots for how the P90 TTFT and TPOT changed as rps increased, for 160m on Xeon Gold 6151 and 1.3b on Xeon Gold 6230 respectively. Each sub-figure has two figures stacked vertically. There are three lines in each of the four figures, representing vLLM, vLLM++, and Sandwich. Lines start flat and curve up. Sandwich sustains the SLO for P90 TTFT and TPOT under a higher rps, while other backends only satisfy P90 TPOT requirement at a lower rps and constantly fail at the P90 TTFT requirement.}
    \label{fig:serve-online}
    \vspace{-4mm}
\end{figure}

\subsection{Serving Benchmark}\label{sec:serve}

\noindent
\textbf{SLOs.}
Since there exist no available SLO settings for these models as far as what we know, we set the SLOs empirically. 
In single sequence serving, time-to-first-token (TTFT) SLO is set to 2200ms and time-per-output-token (TPOT) SLO to 70ms  for the 1.3B model; TTFT is set to 8000ms and TPOT to 240ms for the 8B model.
In batch serving, P90 of TTFTs should be no larger than 200ms and P90 of TPOTs 100ms for the 160M model; P90 TTFT is set to 2000 ms and P90 TPOT to 280 ms for the 1.3B model. We also introduce \textit{SLO scale} to 
vary the SLO latencies. A lower SLO scale corresponds to more stringent latency requirements.

\noindent
\textbf{Single Sequence Serving.}
We conduct extensive serving experiments under twomodel sizes (1.3B, and 8B) on four platforms. 
As shown in Fig.~\ref{fig:main_exp}, \sys{} can sustain 90\% SLO attainment with up to $3.40\times$, and $4.45\times$ more stringent SLOs, 
compared to OpenVINO and vLLM. As for 
token generation throughput and TTFT, \sys{} achieves an average 2.01x higher throughput, shown in Fig.~\ref{fig:main_exp_throu}, and similar TTFT, shown in Fig.~\ref{fig:main_exp_ttft}, across all sample cases compared to the best 
vendor and open-source solutions. 
The tuning time of service configurations used in the experiments 
of \sys{} 
ranges from 78s to 3279s, with an average of 1309s. The kernel tuning time from $[1, seq_{max}]$ ranges from 2022s to 23115s, averaging 10155s.

\noindent
\textbf{Batch Serving.}
We evaluate batch serving on 160M and 1.3B Llama models. Since vLLM supports only BF16 and AVX-512 inference, we conduct experiments on the Xeon 6151 and 6230 processors. 
As we gradually increase the request rate, P90 TPOT and TTFT results are collected. 
Fig.~\ref{fig:serve-online} shows 
that \sys{} can handle a higher request rate (1.84 and 0.92 requests per second)
while meeting the TPOT and TTFT requirements. In contrast, the original vLLM consistently fails to satisfy the TTFT requirement.

\begin{figure}[t]
    \centering
    \includegraphics[width=\linewidth]{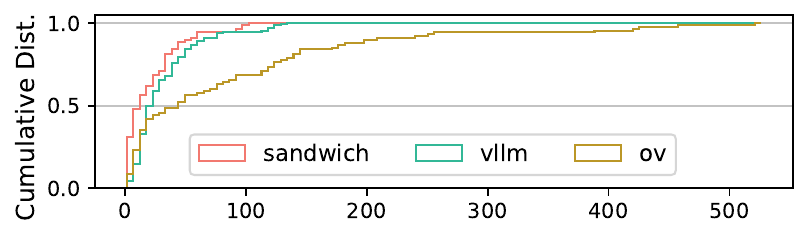}
    
    \Description{A histogram showing distributions of end-to-end request latencies for three backends: sandwich, vllm, and ov. The sandwich distribution has the most number of requests with low latencies, heavily skewed to the left. The vllm distribution is slightly shifted to the right. The ov distribution is shifted to the right further, with a long trailing end of long latencies. }
    \vspace{-3mm}
    \caption{Request latency distribution for Llama3-8B, batch size=1, ShareGPT input, on Xeon 6230.}
    \label{fig:dist_req_time}
\end{figure}

\vspace{1mm}
\noindent
\textbf{Latency Distribution}
Fig.~\ref{fig:dist_req_time} is a CDF plot of the inference latency distribution of requests made to Llama3-8B on Xeon 6320 with batch size 1. \sys{}'s latency distribution shifts significantly toward lower values, indicating that our optimizations greatly reduce inference latency and result in a higher proportion of requests experiencing shorter delays.

\begin{figure}[t]
    \centering
    \vspace{-2mm}
    \includegraphics[width=\linewidth]{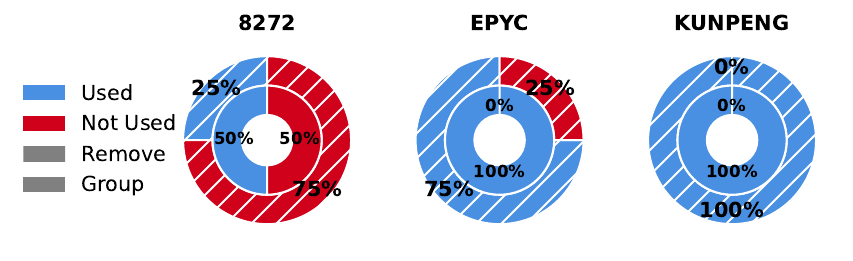}
    \vspace{-10mm}
    \caption{Final distribution of 'Remove' and 'Group' optimizations, where Grouping involves either non-NUMA partitioning or core sub-clustering (latent structure).}
    \label{fig:final_findings}
\end{figure}

\vspace{1mm}
\noindent
\textbf{Findings.}
Shown in Fig.~\ref{fig:final_findings}, we found a dichotomy across platforms with respect to the optimal service configuration composition for 160M, 1.3B, 3B, and 8B models. On the Intel Xeon 8272, core reduction (\texttt{remove}) is used infrequently, and non-NUMA partitioning  (\texttt{group}) does not consistently yield performance gains. For AMD and Kunpeng processors, however, their sub-cluster architecture makes core grouping and reduction scheme essential for optimal performance.

\subsection{Ablation Study}

\begin{table}[t]
\centering
\begin{tabular}{lrr}
\hline
\textbf{Optimizaton} & \textbf{BS=1} & \textbf{BS=8} \\ \hline
vLLM                 & 4.09          & 2.35          \\ \hline
+ communication op   & 13.46         & 3.66          \\
+ service config     & 14.90         & 5.40          \\
+ kernel tuning      & 17.54         & 8.16          \\
+ split k            & 17.09         & 8.78          \\ \hline
\end{tabular}
\caption{Ablation study on different optimizations used in \sys{}. 
Serving Llama 3.2-3B on Xeon 8272CL. The values are throughput (token/s).}
\label{tb:ablation}
\end{table}

\begin{table}[t]
    \centering
    \begin{tabular}{lrr}
    \hline
        $k$ & tuning time (s) & throughput (token/s) \\\hline
        5 & 4,716.86 & 15.46 \\
        10 & 9,609.13 & 15.58 \\
        15 & 13,443.91 & 16.19 \\
        20 & 16,497.87 & 16.42 \\\hline
    \end{tabular}
     \caption{Ablation study on different values of 
     $k$ in selecting the top-$k$ service configurations. 
     Serving single sequences on Llama 1.3b, EPYC 7H12.}
    \label{tab:abl_k}
\end{table}

\begin{table}[t]
    \centering
    \begin{tabular}{lrrr}
    \hline
        $\rho$ & tuning time (s) & TTFT (ms) & throughput (token/s) \\\hline
        5 & 490.99 & 645.07 & 15.38 \\
        10 & 377.92 & 627.14 & 14.86 \\
        15 & 972.93 & 621.76 & 15.51 \\
        20 & 832.30 & 590.22 & 15.48 \\\hline
    \end{tabular}
    \caption{Ablation study on different values of hyperparameter $\rho$ used in \sys{}. Serving single sequences on Llama 1.3b, EPYC 7H12.}
    \label{tab:abl_rho}
\end{table}

We conduct an ablation study 
to demonstrate the latency reduction and 
throughput gains achieved by different techniques in \sys{}. Table \ref{tb:ablation} shows that our communication optimization, service configuration tuning,
and kernel tuning each improve token generation throughput by varying degrees under different batch sizes. Notably, 
split-k (parallelizing the reduction dimension)
does not benefit single-sequence serving, but it improves small-batch (BS=8) throughput. The variation arises from the presence of search space pruning. Employing split-k needs careful tuning. Table \ref{tab:abl_k} shows that as more top-$k$ service configurations from trace autotune are considered in the tensor program generation, higher throughputs are achieved in the LLM serving at the cost of extra tuning time. Table \ref{tab:abl_rho} shows an ablation study of the sliding window size $\rho$ in the tensor program generation. A larger $\rho$ allows consideration of more MKs before the computation slice expands to a large enough size where the MK is stable. A larger $\rho$ yields better prefill performance and also shorter tuning time in some cases, since profiling an inefficient computation slice takes more time. 
\section{Related Works}

\noindent
\textbf{Tree-based Hardware Abstraction.}
Tree-based abstractions~\cite{Fatahalian2006SequoiaPT,Yan2009HierarchicalPT,ghane2019gecko} have been used for describing memory hierarchies of shared memory systems, such that nodes represent memory buffers (e.g., DRAM, caches) shared by its children. Such abstractions are usually used for describing task assignments and memory allocations. \sys{} introduces TopoTree, a mutable memory hierarchy representation, to accelerate the search for latent shared resources and suitable service configurations. 

\vspace{1mm}
\noindent
\textbf{Parallel Tensor Program Optimizations.} The MK-based approach to optimize dynamic-shaped linear algebra programs could be traced back to the HPC era~\cite{Mathur1994MultiplicationOM,Gunnels2001AFO,Gallivan1988ImpactOH}. Manual implementation and experiments have concluded that computing MKs to introduce data locality is essential for optimal performance on systems with a hierarchical memory, and the size of Mks depend on the input shape. \sys{} particularly handles the skewed input shapes (with a dynamic relatively small sequence length dimension and limited parallel potential) that appeared in the LLM prefill phase. 

\vspace{1mm}
\noindent
\textbf{Tensor Program Compilers.} Static-shaped tensor program compilers like AutoTVM/Ansor~\cite{autoTVM,Ansor} utilize an evolutionary algorithm to search a huge optimization space. Dynamic-shaped compilers take the scale-up-and-out approach, proposed by Roller~\cite{Roller} or the cost-model-based approach, proposed by DietCode~\cite{dietcode} and MikPoly~\cite{milkpoly}. \sys{} surpasses the static-shaped approach in tuning speed because of a sophisticated search space and exceeds dynamic-shaped approaches in performance because of its joint consideration of computation slice and polymerization scheme. 



\section{Conclusion}
This paper introduces \sys{}, a compilation framework for efficient CPU-based LLM serving. \sys{} constructs a hardware-centric search space to optimize CPU core utilization and model partition granularity, thereby enhancing serving performance.
\sys{}’s three core innovations—seamless phase-wise plan switching, TopoTree-based substructure-aware core allocation, and fast-start-then-finetune kernel generation—enable it to deliver efficient core utilization plans and generate tensor programs with significantly reduced tuning time. These capabilities allow \sys{} to outperform all existing solutions, unlocking the full potential of CPUs for LLM serving.
Extensive evaluations show that \sys{} delivers significant throughput gains, substantial Goodput improvements, and supports stricter latency SLOs for real-world LLM serving.


\bibliographystyle{ACM-Reference-Format}
\bibliography{paper}

\clearpage
\appendix
\section{Proof of the complexity}\label{sec:transformation_complex}
Our goal is to find $\bigO(\TG(n))$, the upper bound for our search space. Suppose $\bigF(n, k)$ gives the number of all possible ways to partition $n$ subgraphs into groups of size $k$, satisfying our requirements. Now we could define $\TG$ inductively. 

\begin{align}
\TG(1) &= 1 \\
\TG(n) &= \sum_{k \in [2..n]} \TG(\frac{n}{k}) \cdot \bigF(n, k)
\end{align}

Suppose $\smlF(n, k, t)$ gives the number of all possible ways to partition $n$ subgraphs into groups of size $k$, which could tile the original subgraph with a stride of $t$. Now we define $\bigF$ as

$$\bigF(n, k) = \sum_{t \in [1..\lfloor\frac{n}{k}\rfloor]} \smlF(n, k, t)$$

Suppose we have a function $f(n, k, t)$ that givesthe number of possible grouping for $n$ cpus with groups of size $k$ that could tile $n$ using a stride of $t$. We have

$$\smlF(n, k, t) \leq 1$$

Finally, we have
\begin{align}
\bigO(\TG(n)) &= \bigO\left(\sum_{k \in [1..n]} \TG(\frac{n}{k}) \cdot \sum_{t \in [1..\lfloor\frac{n}{k}\rfloor]} 1 \right) \\
&= \bigO\left(\left(\sum_{t \in [1..n]} t \right) \cdot \left( \TG\left(\frac{n}{2}\right) + \TG\left(\frac{n}{3}\right) + \cdots + \TG\left(\frac{n}{n}\right) \right) \right) \\
&= \bigO\left(n \cdot \left(1 + \frac{n}{2} \right) \cdot \left( \TG\left(\frac{n}{3}\right) + \cdots + \TG\left(\frac{n}{n-1}\right) + \TG\left(1\right) \right) \right) \\
&= \bigO\left(n \cdot \left(1 + \frac{n}{2} \right) \cdot \left(1 + \frac{n}{3} \right) \cdots \left(1+\frac{n}{n-1}\right) \cdot 1 \right) \\
&= \bigO\left(\frac{n^n}{(n-1)!}\right)
\end{align}

\subsection{Proof of the Remove Complexity}
We analyze the complexity of the remove transformation in our search space, denoted as $O(S_r(n))$. Let $S(n)$ represent the maximum number of operations needed for a transformation involving $n$ elements, which is characterized by recursive partitioning of the set. We express this relationship as follows:

\begin{equation}
\begin{aligned}
S(n) &= \max_{k_0 \in [2, \ldots, n]} \left( (k_0 - 1) \times S\left(\frac{n}{k_0}\right) \right), \\
S\left(\frac{n}{k}\right) &= \max_{k_1 \in [2, \ldots, \frac{n}{k}]} \left( (k_1 - 1) \times S\left(\frac{n}{k \times k_1}\right) \right), \\
&\vdots
\end{aligned}
\end{equation}

To derive the upper bound, we examine the recursive relationship under the assumption that each division of the set maximizes the number of operations, which occurs when each partition is approximately equal in size. This assumption simplifies the analysis by focusing on the largest contributor to the complexity in each recursive step.

Consider the following cases for the recursive layering, denoted by $L$:

Case $L=2$: The worst-case scenario occurs when the set is divided into $\sqrt{n}$ groups of $\sqrt{n}$ each, resulting in $n$ operations.
Case $L=3$: Extending the argument, if the next layer also divides each group into $\sqrt{n/k}$, the complexity for this layer becomes $\frac{n}{k} \times k = n$, and the optimal choice for $k$ remains $\sqrt{n}$.
Generalizing from these observations, the recursive division continues, reducing the size of the problem in each layer by a factor that is a power of 2 (i.e., $n^{1/2}, n^{1/4}, \ldots, n^{1/2^k}$). Summing over these contributions, the overall complexity can be expressed as:

\begin{equation}
O(S_r(n)) = n \times n^{1/2} \times n^{1/4} \times \ldots \times n^{1/2^k}.
\end{equation}

Each step in the recursion exponentially decreases the size of the problem, and the total complexity converges to $n^2$, since the sum of the exponents in the product approaches 2:

\begin{equation}
1 + \frac{1}{2} + \frac{1}{4} + \ldots + \frac{1}{2^k} \rightarrow 2 \quad \text{as} \quad k \rightarrow \infty.
\end{equation}

Thus, we conclude that the complexity of the remove transformation in our search space is $O(n^2)$.
\begin{table*}[t]
\centering

\begin{tabular}{llllrr}
\toprule
cpu & model & dtype & dataset & graph tune time (s) & kernel tune time (s) \\
\midrule
Xeon Gold 6230 & Felladrin/Llama-160M-Chat-v1 & bfloat16 & shapegpt & 1212 & 7092 \\
Xeon Gold 6230 & Felladrin/Llama-160M-Chat-v1 & bfloat16 & lmsys-chat-1m & 632 & 8673 \\
Xeon Gold 6230 & princeton-nlp/Sheared-LLaMA-1.3B & bfloat16 & shapegpt & 1106 & 7352 \\
Xeon Gold 6230 & princeton-nlp/Sheared-LLaMA-1.3B & bfloat16 & lmsys-chat-1m & 682 & 6032 \\
Xeon Gold 6230 & meta-llama/Meta-Llama-3-8B & bfloat16 & shapegpt & 2797 & 7544 \\
Xeon Gold 6230 & meta-llama/Meta-Llama-3-8B & bfloat16 & lmsys-chat-1m & 1389 & 12747 \\
Xeon Gold 6151 & Felladrin/Llama-160M-Chat-v1 & bfloat16 & shapegpt & 496 & 9218 \\
Xeon Gold 6151 & Felladrin/Llama-160M-Chat-v1 & bfloat16 & lmsys-chat-1m & 728 & 8142 \\
Xeon Gold 6151 & princeton-nlp/Sheared-LLaMA-1.3B & bfloat16 & shapegpt & 1098 & 13680 \\
Xeon Gold 6151 & princeton-nlp/Sheared-LLaMA-1.3B & bfloat16 & lmsys-chat-1m & 142 & 14237 \\
Xeon Gold 6151 & meta-llama/Meta-Llama-3-8B & bfloat16 & shapegpt & 3109 & 18168 \\
Xeon Gold 6151 & meta-llama/Meta-Llama-3-8B & bfloat16 & lmsys-chat-1m & 1202 & 16422 \\
EPYC 7H12 & Felladrin/Llama-160M-Chat-v1 & float32 & shapegpt & 1915 & 2023 \\
EPYC 7H12 & Felladrin/Llama-160M-Chat-v1 & float32 & lmsys-chat-1m & 1439 & 2717 \\
EPYC 7H12 & princeton-nlp/Sheared-LLaMA-1.3B & float32 & shapegpt & 1629 & 10540 \\
EPYC 7H12 & princeton-nlp/Sheared-LLaMA-1.3B & float32 & lmsys-chat-1m & 2157 & 4755 \\
EPYC 7H12 & meta-llama/Meta-Llama-3-8B & float32 & shapegpt & 2664 & 7912 \\
EPYC 7H12 & meta-llama/Meta-Llama-3-8B & float32 & lmsys-chat-1m & 1788 & 11843 \\
Kunpeng 920 & Felladrin/Llama-160M-Chat-v1 & float32 & shapegpt & 262 & 3279 \\
Kunpeng 920 & Felladrin/Llama-160M-Chat-v1 & float32 & lmsys-chat-1m & 79 & 4004 \\
Kunpeng 920 & princeton-nlp/Sheared-LLaMA-1.3B & float32 & shapegpt & 638 & 11410 \\
Kunpeng 920 & princeton-nlp/Sheared-LLaMA-1.3B & float32 & lmsys-chat-1m & 303 & 13386 \\
Kunpeng 920 & meta-llama/Meta-Llama-3-8B & float32 & shapegpt & 3280 & 19446 \\
Kunpeng 920 & meta-llama/Meta-Llama-3-8B & float32 & lmsys-chat-1m & 679 & 23115 \\
\bottomrule
\end{tabular}
\caption{Tuning costs of \sys{} under different scenario}
\end{table*}

\begin{figure}
    \centering
    \includegraphics[width=\linewidth]{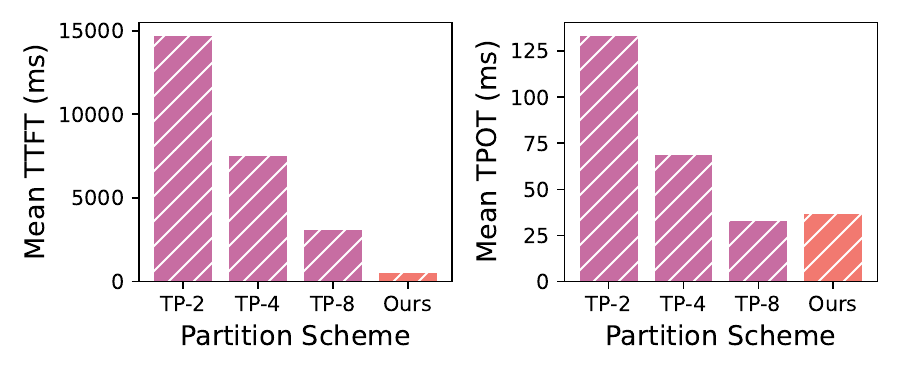}
    \caption{The mean TTFT and TPOT of xFasterTransformers serving Llama-1.3b under different partition schemes compared to our solution. }
    \label{fig:xft_result}
\end{figure}

\section{Other Results}
We present a comparison with xFasterTransformer~\cite{xFasterTransformer} in BF16 on Xeon Gold 6230. As illustrated in Fig.~\ref{fig:xft_result}, \sys{} achieves substantial improvements in TTFT while maintaining compatible TPOT results, demonstrating its robust performance. 

\end{document}